\documentclass[a4paper,12pt]{article}

\usepackage[utf8]{inputenc}
\usepackage{fontawesome5}

\usepackage{amsmath}
\usepackage{amssymb}
\usepackage{amsthm}
\usepackage{bm}
\usepackage{bbm}
\usepackage{mathtools}
\usepackage{stmaryrd}
\usepackage{siunitx}

\usepackage{graphicx}
\usepackage{xcolor}
\usepackage{tikz}
\usetikzlibrary{positioning}

\usepackage{multirow}
\usepackage{array}
\usepackage{booktabs}

\usepackage{natbib}

\usepackage{verbatim}
\usepackage{enumerate}
\usepackage{textcomp}
\usepackage{algorithmic}
\usepackage{algorithm}
\usepackage{float}
\usepackage{authblk}

\usepackage{hyperref}

\usepackage[lmargin=2.4cm,rmargin=2.25cm,bmargin=3cm,tmargin=2.5cm]{geometry}

\newcommand{\mycomment}[1]{}
\newcommand\given[1][]{\:#1\vert\:}

\newcommand{\appropto}{\mathrel{\vcenter{
  \offinterlineskip\halign{\hfil$##$\cr
    \propto\cr\noalign{\kern2pt}\sim\cr\noalign{\kern-2pt}}}}}
\def\bf#1{\mathbf{#1}}
\def\bs#1{\boldsymbol{#1}}

\usepackage{multirow}

\usepackage{fontawesome5}
\usepackage{hyperref}

\makeatletter
\newcommand{\github}[1]{%
   \href{#1}{\faGithubSquare}%
}
\makeatother

\usepackage[utf8]{inputenc}
\usepackage{amsmath}
\usepackage{graphicx}
\usepackage{natbib}
\usepackage{array}

\usepackage{amssymb}
\usepackage{amsthm}
\usepackage{textcomp}
\usepackage{enumerate}
\usepackage{array}
\usepackage{verbatim}
\usepackage{hyperref}
\usepackage{siunitx}

\usepackage{booktabs}
\usepackage{stmaryrd} 
\usepackage{xcolor}
\usepackage{tikz}
\usetikzlibrary{positioning}

\def\bf#1{\mathbf{#1}}
\def\bs#1{\boldsymbol{#1}}

\usepackage{soul} 

\usepackage{algorithmic} 
\usepackage{algorithm}
\usepackage{bbm}
\usepackage{float}
\usepackage{mathtools}
\DeclareMathOperator*{\argmin}{arg\,min} 
\usepackage{printlen}
\uselengthunit{mm}

\usepackage{bm}
\usepackage{ulem} 

\begin{document}

\title{Towards Data-Conditional Simulation for ABC Inference in Stochastic Differential Equations}
\author[1]{Petar Jovanovski}
\author[2]{Andrew Golightly}
\author[1]{Umberto Picchini}
\affil[1]{\small Department of Mathematical Sciences, Chalmers University of Technology and the University of Gothenburg, Sweden.}
\affil[2]{\small Department of Mathematical Sciences, Durham University, UK.}
\date{}

\maketitle
\begin{abstract}We develop a Bayesian inference method for discretely-observed stochastic differential equations (SDEs). Inference is challenging for most SDEs, due to the analytical intractability of the likelihood function. Nevertheless, forward simulation via numerical methods is straightforward, motivating the use of approximate Bayesian computation (ABC). We propose a conditional simulation scheme for SDEs that is based on lookahead strategies for sequential Monte Carlo (SMC) and particle smoothing using backward simulation. This leads to the simulation of trajectories that are consistent with the observed trajectory, thereby increasing the ABC acceptance rate. We additionally employ an invariant neural network, previously developed for Markov processes, to learn the summary statistics function required in ABC. The neural network is incrementally retrained by exploiting an ABC-SMC sampler, which provides new training data at each round. Since the SDEs simulation scheme differs from standard forward simulation, we propose a suitable importance sampling correction, which has the added advantage of guiding the parameters towards regions of high posterior density, especially in the first ABC-SMC round.  Our approach achieves accurate inference and is about three times faster than standard (forward-only) ABC-SMC. We illustrate our method in five simulation studies, including three examples from the Chan-Karaolyi-Longstaff-Sanders SDE family, a stochastic bi-stable model (Schl{\"o}gl) that is notoriously challenging for ABC methods, and a two dimensional biochemical reaction network.
\end{abstract}


\section{Introduction}
Stochastic modelling is an important area of applied mathematics for the study of the evolution of systems driven by random dynamics. Statistical inference for stochastic models is therefore of paramount importance in applied problems where data collected at discrete time instants correspond to observations of a continuous-time stochastic process \citep[see e.g.][]{wilkinson2018stochastic}. 
In this work we focus on Bayesian inference for stochastic differential equations (SDEs) observed at discrete time instants. SDEs \citep{oksendal2013stochastic,sarkka2019applied} are fundamental tools for the modelling of continuous-time experiments that are subject to random dynamics. Application of SDEs are found in many areas including finance \citep{steele2001stochastic}, population dynamics  \citep{panik2017stochastic}, systems biology \citep{wilkinson2018stochastic} and the wider applied sciences \citep{fuchs2013inference}. However, with the exception of very few tractable cases, or the use of exact methods for a relatively small class of intractable diffusions (e.g.  \citealp{beskos2006exact,casella2011exact}), SDEs require simulation-based approaches to conduct inference, since transition densities are unavailable in closed-form and hence the likelihood function cannot be obtained analytically. This prevents straightforward frequentist and Bayesian inference, and as a consequence, has generated much research on approximate methods for parameter inference; see \cite{craigmile2022statistical} for a recent review.
Models having an intractable likelihood are a common problem. For this reason, in the past twenty years, there has been increasing interest in simulation-based inference (SBI) methods (often named ``likelihood-free'' methods). The appeal of SBI methods is that they only require forward-simulation of the model, rather than the evaluation of a potentially complicated expression for the likelihood function, assuming it is available. SBI methods, whose most studied member is arguably approximate Bayesian computation (ABC, \citealp{sisson2018handbook}), are in principle agnostic to the complexity of the model. They are general tools that can be applied as long as enough time and computational resources necessary to forward simulate the model are available. Other SBI methods that follow this logic are, for example, {the bootstrap particle filter when incorporated into} pseudo-marginal methods, such as particle Markov chain Monte Carlo (pMCMC, \citealp{andrieu2009pseudo,andrieu2010particle}), {as the bootstrap particle filter only requires forward simulation from the model}. However, it can be difficult to initialize pMCMC algorithms at suitable parameter values, and this search may require a large number of model simulations (``particles'') to obtain a reasonably mixing chain, when the tested model parameter is outside the bulk of the posterior. The latter is less problematic for ABC algorithms such as those based on sequential Monte Carlo samplers, which is the class of ABC algorithms we examine in this work. Inference returned by ABC methods can be used to initialize pMCMC algorithms, as well as select a covariance matrix for a random walk sampler, as done in  \cite{owen2015scalable}.
While SBI methods are a rapidly expanding class of algorithms (see \citealp{cranmer2020frontier} for a review) that in many cases provide the only viable route to inference for stochastic modelling, they come at a cost. They are computationally demanding, where the bottleneck is the forward simulation of a model (SDEs in our case) conditionally on parameter values $\bs{\theta}$. Here, we consider inference for SDEs when using ABC algorithms, where the proposed SDE trajectories are encouraged to follow the observed data, thus increasing the acceptance rate. In ABC, a ``trajectory'' $\bf{x}$ is a solution  to the SDE (typically obtained via some numerical approximation), simulated conditionally on a given value of the model parameters, and then compared to observed data $\bf{x}^{\rm o}$ via an appropriate distance metric $||\bf{x}-\bf{x}^{\rm o}||$. The procedure is iterated for many proposed values of $\bm{\theta}$ and the parameters generating a small distance $||\bf{x}-\bf{x}^{\rm o}||$ are retained as draws from a posterior distribution approximating the true posterior $\pi(\bm{\theta} \given \bf{x}^{\rm o})$. However, the simulation of $\bf{x}$, which we denote as $\bf{x}\sim p(\bf{x}\given\bm{\theta})$, where $p(\bf{x} \given \bm{\theta})$ is the (unknown) likelihood function of $\bm{\theta}$, is usually ``myopic'' of data $\bf{x}^{\rm o}$. This implies that simulated trajectories can be very distant from data, even when simulated conditionally on plausible parameter values. This behaviour can persist even when observed and simulated data are compared via corresponding summary statistics as $||\bf{s}-\bf{s}^{\rm o}||$. Moreover, summary statistics $\bf{s}=S(\bf{x})$, for some function $S(\cdot)$, are often chosen in an ad-hoc manner, e.g. from domain knowledge or previous literature, though recent semi-automatic approaches have shown their effectiveness \citep{fearnhead2012constructing, chen2020neural, jiang2017learning, forbes2021approximate, wiqvist2019partially}.

In this work we propose, firstly, a conditional simulation scheme for SDE models that utilizes the information provided by the observed data. This leads to simulated trajectories that are ``adapted'' to the observations, and thereby an increase in the acceptance rate for ABC algorithms is obtained. The scheme is based on advancing the ``adapted'' solution paths to the SDE in the time-forward and then in the time-backward direction. In the time-forward direction, a numerical approximation to the ``forward model'' $p(\bf{x} \given \bm{\theta})$ is used to propagate multiple trajectories, and at this stage the trajectories are pointwise weighted according to how consistent they are with the observed data, under the proposed parameter value. In the backward direction, a backward-simulation particle smoother \citep{sarkka2013bayesian} is employed to make use of the weights and construct a single (backward) trajectory $\tilde{\bf{x}}$. We term this forward-backward procedure \textit{data-conditional} simulation, because as opposed to forward simulation, the trajectory is constructed by taking the observed data into account. It is this backward trajectory $\tilde{\bf{x}}$ that is used within ABC. The summary $\tilde{\mathbf{s}} = S(\tilde{\bf{x}})$ of this backward trajectory is no longer distributed according to the forward density $p(\tilde{\mathbf{s}} \given \bs{\theta})$, but rather according to a different density which we denote by $p(\tilde{\mathbf{s}} \given \bs{\theta}, \bf{x}^{\rm o})$. Consequently, the importance weight of the parameter-summary pair takes a different form and includes the intractable summary likelihoods of both the forward and the data-conditional simulator. We circumvent this issue by approximating both intractable likelihoods via their respective synthetic likelihood approximations \citep{wood2010statistical,price2018bayesian}. In our experiments, we show that the additional overhead for computing the approximation of the importance weight via synthetic likelihoods is negligible compared to the gain in acceptance rate. 
Notably, unlike most inference algorithms for SDEs where the simplest numerical scheme is employed (that is, the Euler-Maruyama discretization), our methodology allows the use of higher-order schemes for the time-forward pass, such as the Milstein scheme. 

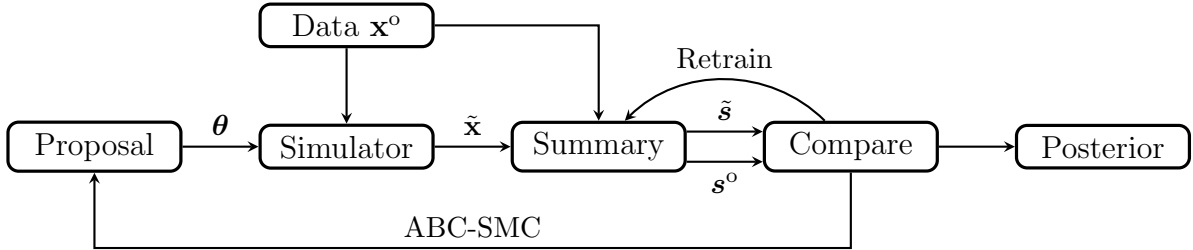
\begin{figure}
    \centering
    \begin{tikzpicture}[>=stealth, 
        node distance=1cm, 
        rectangle_node/.style={rectangle, rounded corners, draw=black, very thick, 
                               align=center, text width=2cm}, 
        small_text/.style={font=\small} 
    ]
    
        \node[rectangle_node] (Proposal) {Proposal};
        \node[rectangle_node, right=of Proposal] (Simulator) {Simulator};
        \node[rectangle_node, right=of Simulator] (Summary) {Summary};
        \node[rectangle_node, right=of Summary] (Compare) {Compare};
        \node[rectangle_node, right=of Compare] (Posterior) {Posterior};
        \node[rectangle_node, above=of Simulator] (Data) {Data $\mathbf{x}^{\mathrm{o}}$}; 
        \draw[->, thick] (Proposal) -- (Simulator) node[midway, above, small_text] {$\boldsymbol{\theta}$};
        \draw[->, thick] (Simulator) -- (Summary) node[midway, above, small_text] {$\tilde{\mathbf{x}}$};
        \draw[->, thick] ([yshift=2mm]Summary.east) -- ([yshift=2mm]Compare.west) node[midway, above, small_text] {$\tilde{\boldsymbol{s}}$}; 
        \draw[->, thick] ([yshift=-2mm]Summary.east) -- ([yshift=-2mm]Compare.west) node[midway, below, small_text] {$\boldsymbol{s}^{\mathrm{o}}$}; 
        \draw[->, thick] (Compare) -- (Posterior);
        \draw[->, thick] (Data) -- (Simulator);
        \draw[->, thick] (Data) -| (Summary);
        \draw[->, thick] (Compare) |- ([yshift=-1cm]Compare.south) -| (Proposal) node[near start, above, small_text] {ABC-SMC};
        \draw[->, thick] (Compare) to[bend right=45] node[midway, above, small_text] {Retrain} (Summary);
    \end{tikzpicture}
    \caption{Diagram of dynamic ABC-SMC with data-conditional simulation. Note the dependence of the simulator on the observed trajectory $\mathbf{x}^{\mathrm{o}}$.}
    \label{fig:diagram}
\end{figure}

Secondly, we construct the summary statistics of univariate SDE models via partially exchangeable networks (PENs, \citealp{wiqvist2019partially}), a recently proposed deep neural network that learns the posterior mean of a Markov process. The data-conditional simulator, together with PEN, is embedded within an ABC sequential Monte Carlo algorithm. In each step, PEN is retrained on newly accepted trajectories, thus sequentially improving the summary statistics. Our approach dramatically speeds up the  convergence to the true posterior, which requires much fewer inference rounds (as we illustrate using Wasserstein distances) and its running time is about 3 times faster  than the ``standard'' (ie using the forward model $p(\bf{x}|\bs{\theta})$) ABC-SMC approach. We find that three rounds of our proposed ABC-SMC scheme typically produces satisfactory results, and note that it could also be used in a ``hybrid'' algorithm, where the output of a few (2-3) rounds of our rapidly converging algorithm could be used to initialize the ``standard'' ABC-SMC approach for the final inference. In Figure \ref{fig:diagram} we present a diagram outlining the structure and flow of our proposed inference pipeline.

The paper is structured as follows. In Sections \ref{sec:abc}-\ref{sec:abc-smc} we give some necessary background details of ABC methodology and in particular ABC-SMC. Section \ref{sec:SDEs} introduces stochastic differential equations and data-augmentation considerations.  Section \ref{sec:backward-sims} introduces our backward-smoothing methodology. Section \ref{sec:synthetic-like} constructs the importance sampling weights for ABC-SMC. Section \ref{sec:summaries} gives the neural network methodology used to learn summary statistics for ABC that are suitable for SDEs. 
Finally all the previous concepts are combined in Section \ref{sec:abc-smc with bs} to give the algorithm for ABC-SMC with data-conditional simulation. Section \ref{sec:examples} contains simulation studies which illustrate the proposed methodology. A Supplementary Material section includes further simulations and methodology. Accompanying code is available on GitHub\footnote{\url{https://github.com/perojov/DataConditionalABC}}.
\subsection{Approximate Bayesian Computation}\label{sec:abc}
Let $p(\bf{x} \given \bs{\theta})$ be the likelihood function of $\bs{\theta}$ corresponding to a given probabilistic model, and let $\pi(\bm{\theta})$ be the prior density ascribed to $\bm{\theta}$. The posterior density of $\bs{\theta}$ given the observed data $\bf{x}^{\rm o}$ is
\begin{equation}
    \pi(\bs{\theta} \given \bf{x}^{\rm o}) \propto p(\bf{x}^{\rm o} \given \bs{\theta})\pi(\bs{\theta}). \label{eq:bayes-rule}
\end{equation}
With the exception of a small subset of SDE models, the likelihood function $p(\bf{x}^{\rm o} \given \bs{\theta})$ cannot be obtained in closed form, and this hinders the direct application of commonly used sampling algorithms for \eqref{eq:bayes-rule}, such as Markov chain Monte Carlo (MCMC).
Approximate Bayesian computation (ABC) bypasses the inability to evaluate the likelihood function analytically, by the ability to forward-simulate from it using a \textit{model simulator}. Here, the model simulator is the SDE solver (up to a discretization error associated with the solver used) which produces a solution $\bf{x}\sim p(\bf{x}|\bm{\theta})$ conditional on some value of $\bm{\theta}$. Using multiple calls from the model simulator, ABC samples parameters from an approximate posterior $\pi_\epsilon(\bm{\theta} \given \mathbf{x}^{\rm o})$, rather than the exact posterior in (\ref{eq:bayes-rule}). Typically, as motivated in detail in section \ref{sec:summaries}, it is necessary to first summarize the data with low-dimensional summary statistics $S(\cdot)$, and sample from $\pi_\epsilon(\bs{\theta} \given S(\mathbf{x}^{\rm o}))$. In this case the ABC posterior becomes
\begin{equation}
    \pi_\epsilon(\bs{\theta} \given S(\mathbf{x}^{\rm o}))  \propto \pi(\bs{\theta}) \int \mathbbm{1}( \Vert S(\mathbf{x}) - S(\mathbf{x}^{\rm o}) \Vert \leq \epsilon) p(S(\mathbf{x}) \given \bs{\theta}) \,  {\rm d}\mathbf{x}, \label{eq:abs-posterior}
\end{equation}
where $||\cdot||$ is an appropriate distance metric, $\mathbbm{1}( \Vert S(\mathbf{x}) - S(\mathbf{x}^{\rm o}) \Vert \leq \epsilon)$ is the indicator function, and $\epsilon>0$ a tolerance value determining the accuracy of the approximation. 
The likelihood of the summary statistics $p(S(\mathbf{x}) \given \bs{\theta})$ is implicitly defined, namely samples from the latter are obtained by first sampling (simulating) $\bf{x} \sim p(\bf{x} \given \bs{\theta})$ and then computing $S(\mathbf{x})$.  
Notable properties of \eqref{eq:abs-posterior} are that if $S(\cdot)$ is highly informative for $\bm{\theta}$, then $\pi_\epsilon(\bs{\theta} \given S(\mathbf{x}^{\rm o}))\simeq \pi_\epsilon(\bs{\theta} \given \mathbf{x}^{\rm o})$ where equality only holds when $S(\cdot)$ is a sufficient statistic (which in practice is never assumed to be the case). Additionally, when $\epsilon\rightarrow 0$, $\pi_\epsilon(\bs{\theta} \given S(\mathbf{x}^{\rm o}))\rightarrow \pi(\bs{\theta} \given S(\mathbf{x}^{\rm o}))$. Therefore, using a small $\epsilon$ and an ``informative'' $S(\cdot)$, ABC may produce a reasonable approximation to the true posterior $\pi(\bs{\theta} \given \mathbf{x}^{\rm o})$. However, finding an appropriate $\epsilon$ is an exercise in balancing the increasing computational effort when $\epsilon$ is reduced (this in turn increases the rejection rate of the proposed parameters), against more accurate posterior inference (provided enough accepted parameter draws have been collected). As a shorthand, we will denote the summary statistics of the observations by $\mathbf{s}^{\rm o} = S(\mathbf{x}^{\rm o})$.

The basic ABC algorithm for producing samples from \eqref{eq:abs-posterior} is the ABC with rejection-sampling (ABC-RS) algorithm \citep{tavare1997inferring, pritchard1999population}, which can be described in three steps. Suppose we have an importance density $g(\bs{\theta})$ that is easy to sample from and has the same support as the prior. Instead of proposing from $\pi(\bs{\theta})$ as is typically done, (i) propose  $\bs{\theta} \sim g(\bs{\theta})$, (ii) simulate (implicitly) the summary statistic $\bf{s} \sim p( \bf{s} \given \bs{\theta})$, and (iii) accept $\bs{\theta}$ as a sample from the ABC posterior with  probability 
\begin{equation}
    \frac{\mathbbm{1}( \Vert \mathbf{s} - \mathbf{s}^{\rm o} \Vert \leq \epsilon)\pi(\bs{\theta})}{C g(\bs{\theta})}, \quad \text{where } C = \max_{\bs{\theta} \in \bs{\Theta}} \frac{\pi(\bs{\theta})}{ g(\bs{\theta})}. \label{eq:accept-probability}
\end{equation}
Steps (i)-(iii) are iterated until a desired number $M$ of parameter draws have been accepted. Since in steps (i) and (ii) both the parameter and the summary statistic are sampled, the actual sampling density of this algorithm is the joint density $g(\bs{\theta}, \bf{s}) = p(\bf{s} \given \bs{\theta}) g(\bs{\theta})$, and the target is the joint ABC posterior $\pi_\epsilon(\bs{\theta}, \bf{s} \given \mathbf{s}^{\rm o})$ of the parameter $\bs{\theta}$ and the summary statistic $\bf{s}$, given by $\pi_\epsilon(\bs{\theta}, \bf{s} \given \mathbf{s}^{\rm o}) \propto \mathbbm{1}( \Vert \mathbf{s} - \mathbf{s}^{\rm o} \Vert \leq \epsilon) p(\mathbf{s} \given \bs{\theta}) \pi(\bs{\theta})$. The sampled summary statistic can be safely discarded since marginalizing $\pi_\epsilon(\bs{\theta}, \bf{s} \given \mathbf{s}^{\rm o})$ over $\mathbf{s}$ gives $\pi_\epsilon(\bs{\theta} \given \mathbf{s}^{\rm o}) = \int \pi_\epsilon(\bs{\theta}, \bf{s} \given \mathbf{s}^{\rm o}) \,  {\rm d}\mathbf{s}$, the ABC posterior \eqref{eq:abs-posterior}. The acceptance probability of ABC-RS is then given by 
\begin{equation}
    \frac{\pi_\epsilon(\bs{\theta}, \mathbf{s} \given \mathbf{s}^{\rm o})}{Cg(\bs{\theta}, \mathbf{s})} \propto \frac{\mathbbm{1}( \Vert \mathbf{s} - \mathbf{s}^{\rm o} \Vert \leq \epsilon) p(\mathbf{s} \given \bs{\theta}) \pi(\bs{\theta})}{C p(\mathbf{s} \given \bs{\theta}) g(\bs{\theta})} = \frac{\mathbbm{1}( \Vert \mathbf{s} - \mathbf{s}^{\rm o} \Vert \leq \epsilon) \pi(\bs{\theta})}{C g(\bs{\theta})}, \label{eq:complete-acc-prob}
\end{equation}
with $C$ defined as in \eqref{eq:accept-probability}. 
The choice of the marginal importance density $g(\bs{\theta})$ is crucial to the efficiency of the algorithm. For example, if $g(\bs{\theta})$ is much more diffuse compared to the posterior density, most proposed parameter draws will be rejected, and if $g(\bs{\theta})$ is too concentrated, the tails of the posterior will not be thoroughly explored. To remedy the inefficiencies of ABC-RS, many modifications have been proposed, such as likelihood-free MCMC \citep{marjoram2003markov, wegmann2009efficient}, \cite[Chapter~12]{brooks2011handbook}, \cite[Chapter~9]{sisson2018handbook}  and likelihood-free sequential Monte Carlo \citep{del2006sequential, sisson2007sequential, peters2012sequential, toni2009approximate, beaumont2009adaptive, del2012adaptive,picchini2022guided}. In the latter family of algorithms, a population of parameter samples (commonly referred to as particles) are propagated along a sequence of $T$ ABC posterior distributions with decreasing acceptance thresholds $\epsilon_1 > \epsilon_2 > \ldots > \epsilon_T$. With each step in the sequence, the ABC posterior $\pi_{\epsilon_t}(\bs{\theta} \given \mathbf{s}^{\rm o})$ becomes less diffuse, and is used to generate parameter particles in the subsequent step. This has the effect of sampling parameter particles from regions of increasingly higher posterior density, by sequentially traversing the $T$ intermediate distributions.

\subsection{Approximate Bayesian Computation - Sequential Monte Carlo}\label{sec:abc-smc}

Consider ABC implemented via a sequential Monte Carlo sampler, denoted  ABC-SMC. The initialization works along the same lines as ABC-RS, however, rather than computing the normalizing constant as in \eqref{eq:complete-acc-prob}, an importance weight $w(\bs{\theta}, \bf{s}) = \mathbbm{1}( \Vert \mathbf{s} - \mathbf{s}^{\rm o} \Vert \leq \epsilon) \pi(\bs{\theta}) / g(\bs{\theta})$
is assigned to each sample $(\bs{\theta}, \bf{s}) \sim g(\bs{\theta}, \bf{s})$. The ABC posterior $\pi_{\epsilon_1}$ at this initial step is constructed from a set of $M$ weighted samples $(\bs{\theta}_1^{1:M}, W_1^{1:M})$ as $\pi_{\epsilon_1}(\bs{\theta} \given \mathbf{s}^{\rm o}) = \sum_{j = 1}^M W_1^j \mathcal{K}(\bs{\theta}_{t}^i \given \bs{\theta}_{t - 1}^j)$ \citep{del2006sequential}. Here, $\mathcal{K}(\cdot \given \cdot)$ serves as a transition kernel to move the particles into regions of (potentially) high posterior density, as well as increase the particle diversity in the population. In the subsequent steps, rather than proposing parameters from the initial importance density $g(\bs{\theta})$, they are proposed from a perturbation of the ABC posterior approximated in the previous step. More precisely, at step $t$, a particle is sampled from the previous population $\bs{\theta}^{1:M}_{t - 1}$ with probabilities $W_{t - 1}^{1:M}$, and then perturbed according to the transition kernel. Since the parameters are sampled from the previous ABC posterior, the importance weights for the new particle population $\bs{\theta}_{t}^{1:M}$ at this step take the form $W_t^i \propto \pi(\bs{\theta}_{t}^i)/ \sum_{j = 1}^M W_{t - 1}^j \mathcal{K}(\bs{\theta}_{t}^i \given \bs{\theta}_{t - 1}^j)$. A typical choice for the transition kernel is the Gaussian density  $\mathcal{K}(\bs{\theta}_{t}^i \given \bs{\theta}_{t - 1}^j) = \mathcal{N}(\bs{\theta}_{t}^i \given \bs{\theta}_{t - 1}^j, \bs{\Sigma}_t)$ \citep{beaumont2009adaptive, toni2009approximate} (here $\mathcal{N}(\bf{x} \given \bs{\mu}, \bs{\Sigma})$ denotes the multivariate normal density evaluated at $\bf{x}$ with mean $\bs{\mu}$ and covariance $\bs{\Sigma}$), where $\bs{\Sigma}$ can be specified in several ways. Algorithm \ref{algo:abs-smc} outlines ABC-SMC with $\bs{\Sigma}_t$ chosen to be twice the (weighted) covariance-matrix of the current particles population (as in \citealp{beaumont2009adaptive,filippi2013optimality}). We note that novel, efficient parameter proposals for ABC-SMC have been recently introduced in \cite{picchini2022guided}, though we will not use those in the present work. Finally, notice that the sequence of thresholds $\epsilon_1 > \ldots > \epsilon_T$ does not need to be fixed as an input to the algorithm, but can be dynamically determined at runtime, typically as a percentile of simulated distances.
\begin{algorithm}

    \caption{ABC-SMC ($\epsilon_1 > \ldots > \epsilon_T)$}\label{algo:abs-smc}
    \begin{algorithmic}[1]
        \FOR{$i = 1$ to $M$}
            \WHILE{parameter not accepted}
                \STATE Sample parameter $\boldsymbol{\theta}^i_1 \sim g(\boldsymbol{\theta})$ and summary $\mathbf{s}^i_1 \sim p(\mathbf{s} \given \boldsymbol{\theta}^i)$.
                 \IF{$\Vert \mathbf{s}^i_1 - \mathbf{s}^{\rm o} \Vert \leq \epsilon_1$} 
                            \STATE Accept $\boldsymbol{\theta}^i_1$ and compute $W_1^i = \pi(\boldsymbol{\theta}^i_1) / g(\boldsymbol{\theta}^i_1)$.
                        \ENDIF
            \ENDWHILE
        \ENDFOR
        \STATE Normalize $W_1^{1:M}$.
        \FOR{$t = 2$ to $T$}
            \STATE Compute particle covariance $\mathbf{\Sigma}_t = 2 \times {\rm Cov}((\bs{\theta}^{1:M}_{t - 1}, W_{t - 1}^{1:M}))$.
            \FOR{$i = 1$ to $M$}
                \WHILE{parameter not accepted}
                    \STATE Sample $\bs{\theta}^*$ from $\bs{\theta}_{t - 1}^{1:M}$ with probabilities $W_{t - 1}^{1:M}$.
                    \STATE Sample $\bs{\theta}^{i}_t \sim \mathcal{N}(\bs{\theta}^*, \mathbf{\Sigma}_t)$ and simulate $\mathbf{s}^i \sim p(\mathbf{s} \given \boldsymbol{\theta}^i_t)$.
                    \IF{$\Vert \mathbf{s}^i - \mathbf{s}^{\rm o} \Vert < \epsilon_t$} 
                        \STATE Accept $\boldsymbol{\theta}^i_t$ and compute $W_t^i = \pi(\bs{\theta}_{t}^i) / \sum_{j = 1}^M W_{t - 1}^j \mathcal{N}(\bs{\theta}_{t}^i \given \bs{\theta}_{t - 1}^j, \mathbf{\Sigma}_t)$.
                    \ENDIF
                \ENDWHILE
            \ENDFOR
            \STATE Normalize $W_t^{1:M}$.
        \ENDFOR
        \STATE \textbf{Output:} Weighted sample $(\boldsymbol{\theta}^{1:M}_T, W_{T}^{1:M})$ of the ABC posterior density.
    \end{algorithmic}
\end{algorithm}

\section{Inferential problem and numerical simulation of SDEs}\label{sec:SDEs}

Consider the time-homogeneous It\^{o} diffusion $(\mathbf{X}_t)_{t \geq 0}$ satisfying the SDE
\begin{equation}
    \begin{cases}
        {\rm d}\mathbf{X}_t = \boldsymbol{\mu}(\mathbf{X}_t, \boldsymbol{\theta}) {\rm d}t + \boldsymbol{\sigma}(\mathbf{X}_t, \boldsymbol{\theta}) {\rm d} \mathbf{B}_t, & \text{if } t > 0, \\
        \mathbf{X}_0 = \mathbf{x}_0, & \text{if } t = 0.
    \end{cases}
    \label{eq:sde}
\end{equation}
with state space $\mathcal{X} \subseteq \mathbb{R}^d$, constant initial condition $\mathbf{x}_0 \in \mathcal{X}$, $p$-dimensional model parameter $\boldsymbol{\theta} \in \boldsymbol{\Theta} \subseteq \mathbb{R}^p$ and $d$-dimensional standard Brownian motion $(\mathbf{B}_t)_{t \geq 0}$. The drift function $\boldsymbol{\mu}: \mathcal{X} \times \boldsymbol{\Theta} \to \mathbb{R}^d$ and diffusion coefficient $\boldsymbol{\sigma}: \mathcal{X} \times \boldsymbol{\Theta} \to \mathbb{R}^{d \times d}$ are assumed to be known in parametric form, and assumed to be sufficiently regular to ensure the existence and uniqueness of a solution for (\ref{eq:sde}). We denote the transition density of the process $(\mathbf{X}_t)_{t \geq 0}$ by $p(\bf{x}_t \given \bf{x}_s, \bs{\theta})$ for times $0\leq s<t$. 

The goal of our work is to perform statistical inference on the parameter $\boldsymbol{\theta}$, based on discrete-time observations of the diffusion at time instants $t_1 < t_2 < \ldots < t_n$, {observed exactly (ie without additional measurement error)}. Due to the Markov property, the likelihood of $\boldsymbol{\theta}$ factorizes into the product of mutually independent transition densities. More precisely, for observation $\mathbf{x}^{\rm o} = (\mathbf{x}^{\rm o}_0, \mathbf{x}^{\rm o}_{1}, \ldots, \mathbf{x}^{\rm o}_{n})$, with a non-random initial state $\bf{x}^{\rm o}_0 = \bf{x}_0$, the likelihood is given by
\begin{equation}
    p(\mathbf{x}^{\rm o}_{1}, \ldots, \mathbf{x}^{\rm o}_{n} \given \boldsymbol{\theta}) = \prod_{i=1}^n p(\mathbf{x}^{\rm o}_{i} \given \mathbf{x}^{\rm o}_{{i - 1}}, \boldsymbol{\theta}).
\end{equation}
If the transition densities $p(\mathbf{x}^{\rm o}_{i} \given \mathbf{x}^{\rm o}_{{i - 1}}, \boldsymbol{\theta})$ are analytically tractable, statistical inference may proceed via maximum likelihood estimation, or via a Bayesian approach by computing the posterior distribution of $\bs{\theta}$ given the observation $\bf{x}^{\rm o}$. However, the transition densities are tractable only for a handful of specific SDE models, and hence the likelihood of $\bs{\theta}$ cannot in general be evaluated. 
There exist several different approaches to deal with the problem of intractable likelihoods, see e.g. \citep{ iacus2008simulation,fuchs2013inference, craigmile2022statistical} for a survey. 
In this work we focus on methods that utilize ``data augmentation'', implying the need to numerically discretize the solution of the SDE on a finer time-scale than the observational grid $(t_1,t_2,...,t_n)$. {We assume a regular observation grid with $t_i - t_{i - 1} = \Delta$ for simplicity in notation. However, our method is also applicable to grids with irregular spacing, though special attention will need to be paid in these cases, see Section 6 in the Supplementary Material.} 
Approximate solutions can be simulated using numerical methods that are based on the discrete-time approximation of the continuous process. Additionally, the quality of the ABC approximation to the posterior depends on the accuracy of the simulator. Sometimes exact simulation is possible (e.g.  \citealp{beskos2006exact,casella2011exact}), but in the case where the simulator is a numerical discretization method, e.g. Euler-Maruyama (EM), the accuracy will be determined by the fineness of the grid. In applications, the inter-observation times are often large, and direct simulation on that same scale will lead to bias. Bayesian data augmentation methods \citep{fuchs2013inference, van2017bayesian, papaspiliopoulos2013data, golightly2022augmented} introduce missing data between the observations such that the union of the observation and the missing data forms a dataset on a finer scale. Motivated by these approaches, we introduce a finer discretization of the interval $(t_0, t_1, \ldots, t_n)$ by $(\tau_0, \tau_1, \ldots, \tau_{A}, \ldots \tau_{nA})$, where $\tau_{iA} = t_i$, and such that $\tau_k - \tau_{k - 1} = h = \Delta / A$, where $A \geq 2$ is the number of subintervals between two observational time instants. We denote by $N \equiv nA$ the total number of elements in the finer discretization, and for the discrete values of $(\bf{X}_t)_{t \geq 0}$ we adopt the shorthand $\bf{x}_{iA + k} \equiv \bf{x}_{\tau_{iA + k}}$. Similarly for the observed trajectory, we adopt the shorthand $\bf{x}^{\rm o}_{i} \equiv \bf{x}^{\rm o}_{t_i}$.
The EM method is the most commonly used numerical scheme for SDEs: given the aforementioned finer partition, EM produces the discretization \begin{equation}
    \mathbf{X}_{{i + 1}} = \mathbf{X}_{{i}} + \boldsymbol{\mu}(\mathbf{X}_{i}, \boldsymbol{\theta}) h + \boldsymbol{\sigma}(\mathbf{X}_{i}, \boldsymbol{\theta}) (\mathbf{B}_{{i + 1}} - \mathbf{B}_{i}),  
\end{equation}
for $i = 0, \ldots, N - 1$. By the properties of the Brownian motion, the transition density $p(\mathbf{x}_{{i + 1}} \given \mathbf{x}_{i}, \bs{\theta})$ induced by the EM scheme is multivariate Gaussian with mean $\mathbf{x}_{i} + \boldsymbol{\mu}(\mathbf{x}_{i}, \boldsymbol{\theta}) h$ and covariance matrix $\boldsymbol{\sigma}(\mathbf{x}_{i}, \boldsymbol{\theta}) \boldsymbol{\sigma}^T(\mathbf{x}_{i}, \boldsymbol{\theta}) h$. 

\begin{figure}
    \centering\includegraphics[scale=1]{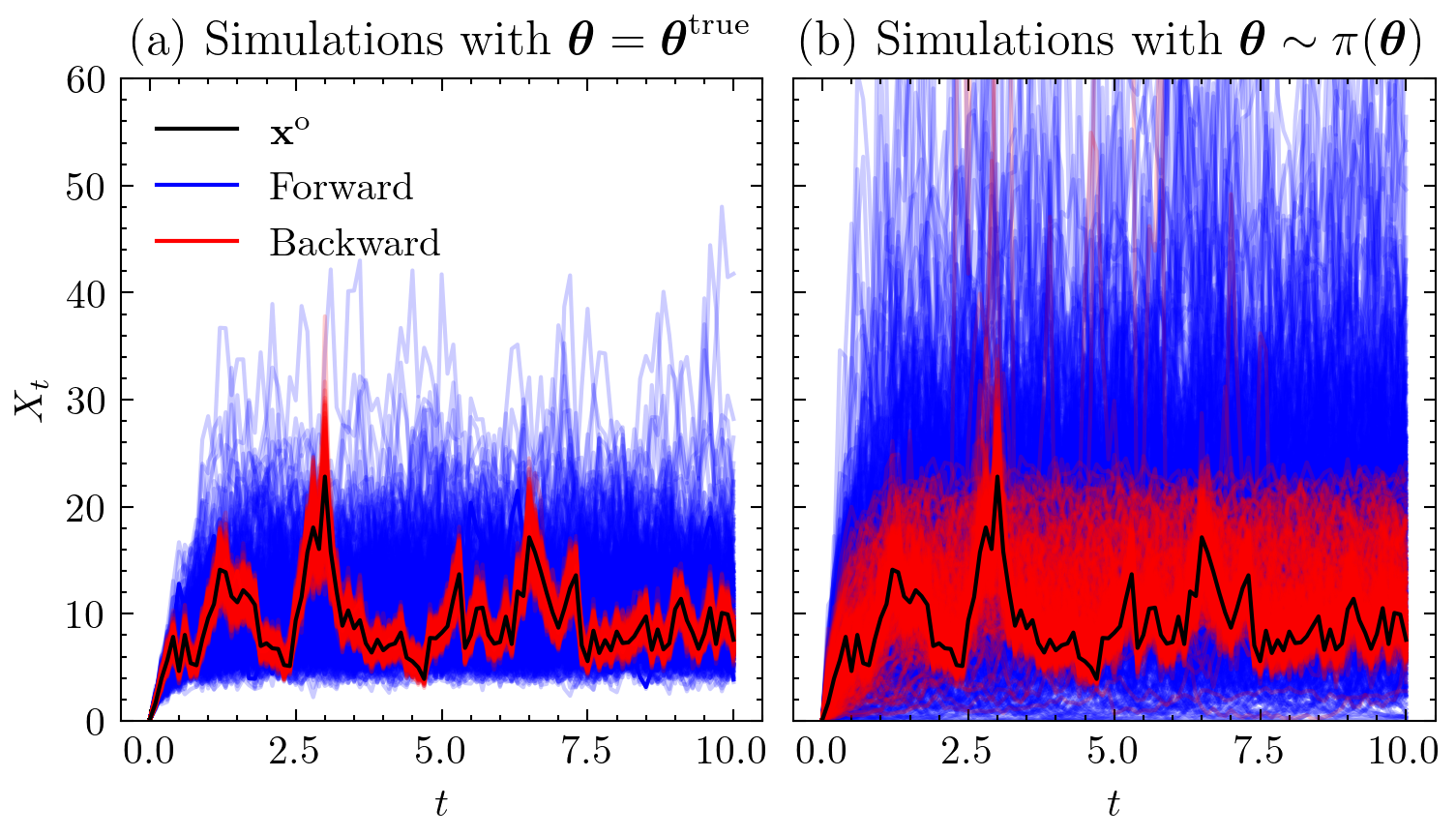}
    \caption{Illustration of simulated trajectories from both the forward and the data-conditional simulator with $P = 30$ particles for the CKLS SDE ${\rm d}X_t = \beta (\alpha - X_t) {\rm d}t + \sigma X_t^{\gamma} {\rm d}B_t$. The observation is shown in black, the forward trajectories in blue, and the backward trajectories in red. In (a) 500 trajectories (per simulator) are shown: these were simulated conditionally on the true parameter $\bs{\theta}^{\rm true}$. In (b) 500  trajectories (per simulator) are shown: these were simulated from the prior-predictive distribution, with prior $\pi(\bs{\theta}) = \mathcal{U}(5, 30)\mathcal{U}(0, 3)\mathcal{U}(0.5, 2)\mathcal{U}(0.6, 1)$. This prior is chosen only for illustration, and it is different from the prior used for inference.}
    \label{fig:ckls-trajectories-true-and-pps}
\end{figure}

In this paper we will consider two ways to simulate a trajectory: the first is ``myopic'' forward simulation, meaning that the trajectory is generated from $p(\bf{x} \given \boldsymbol{\theta})$ and is therefore not conditioned on the observation (this is the standard approach in ABC, e.g. \citealp{picchini2014inference}). The second way is one of our contributions which we term \textit{data-conditional} simulation, where the sample path is generated from $p( \bf{x} \given \boldsymbol{\theta}, \bf{x}^{\rm o})$, and thereby conditional on the observed data. In section \ref{sec:backward-sims} we will construct this scheme which is based on a forward/backward  idea. The scheme possesses a ``lookahead'' mechanism for which the distribution of the intermediate points depends on the successive observations. In the forward direction, multiple trajectories are forward propagated and weighted, and then in the backward direction a single trajectory is simulated using a backward-simulation particle smoother. For an illustration of the difference between these two approaches, see Figure \ref{fig:ckls-trajectories-true-and-pps}. There, for the CKLS SDE that we consider extensively in Section \ref{sec:examples}, we show that, due to the specific diffusion term, the trajectories produced from the forward model can vary substantially, even when simulating conditionally on the true parameter (Figure \ref{fig:ckls-trajectories-true-and-pps}(a)). However using our forward-backward data-conditional approach, simulated trajectories are much more consistent with the observed data. Data-conditional simulation is illustrated in Section \ref{sec:backward-sims}. 

\section{Data-conditional simulation}\label{sec:backward-sims}
Backward simulation can be implemented via a sequential Monte Carlo (SMC) scheme that is based on processing the observation in the forward and then in the backward direction. In the forward direction, a numerical scheme  is used to propagate multiple particles, which represent numerical solutions to the underlying SDE model. At this stage the particles are suitably weighted according to how close they are to the observed data $\bf{x}^{\rm o}$. In the backward direction, a backward-simulation particle smoother (BSPS) is applied on the particle system that is obtained from the forward direction, in order to construct a single (backward) trajectory. This backward trajectory is a sample from the conditional distribution $p(\bf{x} \given \bs{\theta}, \bf{x}^{\rm o})$.
The development of the forward direction of our data-conditional simulator borrows ideas from the literature of simulation methods for diffusion bridges using SMC, such as \cite{del2005genealogical, lin2010generating} and in particular, the bridge particle filter (BPF) of \cite{del2015sequential} (see also \citealp{golightly2015bayesian}). In the BPF, the simulation of a diffusion bridge $\bf{x}_{{iA}}, \ldots, \bf{x}_{{(i + 1)A}}$ (recall that $A$ is the number of subintervals) between two consecutive observations $(\bf{x}^{\rm o}_{i}, \bf{x}^{\rm o}_{{i + 1}})$ relies on a suitably chosen particle weighting scheme, such that a set of $P$ particles are guided from $\bf{x}^{\rm o}_{i}$  to $\bf{x}^{\rm o}_{{i + 1}}$. Particles are simulated forward in time using only $p(\bf{x}_{k} \given \bf{x}_{{k - 1}}, \bs{\theta})$ for $iA \leq k \leq (i + 1)A$, that is, the transition density arising from a numerical scheme such as Euler-Maruyama, or if desired, a higher order scheme (e.g. the Milstein scheme in $d=1$). 
Even though the particles are propagated by the forward density, they are weighted in such a way that, by looking ahead to the next observation point $\bf{x}^{\rm o}_{{i + 1}}$, those particle trajectories that are not consistent with $\bf{x}^{\rm o}_{{i + 1}}$ will be given small weights and pruned out with a resampling step which occurs at every intermediate time point $\tau_k$ (unlike in standard particle filtering where resampling occurs at observation time points only). However, the BPF  will result in a particle system $(\bf{x}^{1:P}_{1:N}, \omega^{1:P}_{1:N})$ for which $(\bf{x}^j_{1:N})$ is, at the observation time instants, exactly equal to the observed data, i.e. $\bf{x}^j_{{t_i}} \equiv \bf{x}^{\rm o}_{i}$ for $i = 0, \ldots, n$. The reason is that for every subinterval $[t_i, t_{i + 1}]$ the states $\bf{x}_{{iA + 1}}, \ldots, \bf{x}_{{(i + 1)A - 1}}$ are randomly generated, whereas $\bf{x}_{{iA}} = \bf{x}^{\rm o}_{{i}}$, $\bf{x}_{{(i + 1)A}} = \bf{x}^{\rm o}_{{i + 1}}$ is fixed. The consequence is that, in the backward pass, the backward-simulation particle smoother will construct a backward trajectory that is exactly equal to $\bf{x}^{\rm o}$ (at the observation time instants), regardless of the value of $\bs{\theta}$. Hence, the event $\{\Vert \mathbf{s} - \mathbf{s}^{\rm o} \Vert = 0\}$ in \eqref{eq:complete-acc-prob} will occur with probability 1 for any $\bs{\theta}$, which is unwanted. Rather, we require acceptance for suitable values of $\bs{\theta}$, i.e. those that are representative of the posterior density. To this end, we will change the BPF as follows. 
In our adaptation we make two modifications. The first modification omits the resampling step in the BPF, in order for the particles (which represent numerical solutions to the underlying SDE model) to preserve their original sampling distribution $p(\bf{x}_{k} \given \bf{x}_{{k - 1}}, \bs{\theta})$. This will be of fundamental importance when we embed the data-conditional simulator within ABC-SMC, as well as in assembling the training data for the partially exchangeable neural network. Although we omit the resampling step, the particle weights are still computed, and they will be utilized once the forward phase is completed. The second modification is that we perform one more forward propagation step, {from time $\tau_{(i + 1)A - 1}$ to $\tau_{(i + 1)A}$ for every $i = 0, \ldots, n - 1$, namely, the states $\bf{x}_{(i + 1)A}$ are random, as opposed to be deterministically set to $\bf{x}^{\rm o}_{i + 1}$ as in the BPF of \cite{del2015sequential}.} 

The resulting algorithm reduces to sequential importance sampling (SIS) and will yield forward (myopic) trajectories that are weighted with respect to the observations. In order to utilize the weights, once the trajectories have been simulated up to time $t_n$, we complement SIS with a second recursion that evolves backward in time. In the backward pass, the (backward) trajectory is produced by firstly simulating $\tilde{\bf{x}}_{N}$, then $\tilde{\bf{x}}_{{N - 1}}$, etc. until a complete trajectory $\tilde{\bf{x}}_{{0}}, \ldots, \tilde{\bf{x}}_{{N}}$ is generated. This forward-backward idea will allow us to develop a simulator that will generate trajectories which are consistent with the observation $\mathbf{x}^{\rm o}$. The procedure is detailed in next sections.

\subsection{Lookahead sequential importance sampling}
\begin{figure}
    \centering
    \includegraphics[scale=0.67]{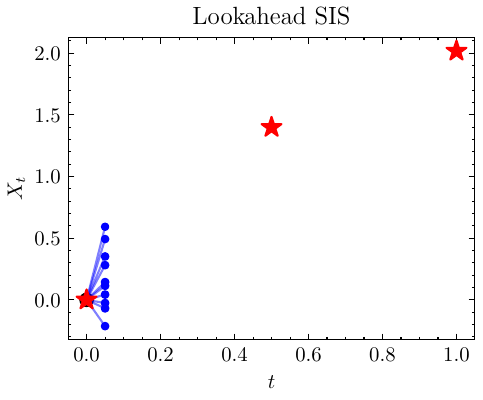}
    \includegraphics[scale=0.67]{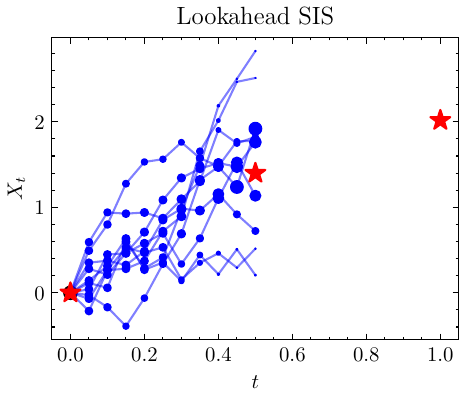}
    \includegraphics[scale=0.67]{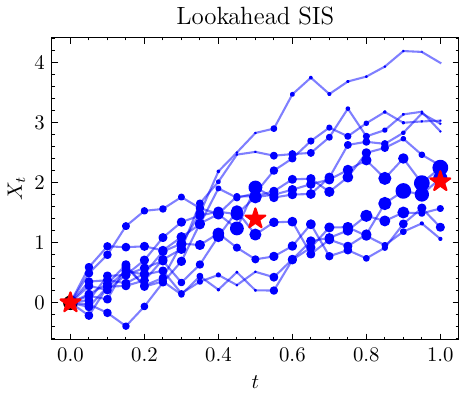}
    \caption{Evolution of a particle system obtained from the Lookahead SIS algorithm. Observations (red stars) and particles (blue). The size of the particles weights are depicted by the sizes of the circles. Left: the particles are initialized at the initial state, and a single forward step is performed. Middle: the particles are simulated up to time $t_1$ and are weighted according to the observed state $x^{\rm o}_{1}$. The closer the particle is to the observation, the larger its weight becomes. At time $t_1$, the particles start to look towards the subsequent observation $x^{\rm o}_{2}$, and therefore are weighed in a different way, as seen in the right panel.}
    \label{fig:lookahead-sis}
\end{figure}

We will begin by deriving the lookahead SIS algorithm for the first observation interval $[0, t_1]$, before considering the remaining intervals. As motivated above, the time interval $[0, t_1]$ is partitioned into $A$ subintervals $0 = \tau_0 < \tau_1 < \ldots < \tau_{A} = t_1$, and the diffusion takes values $\bf{x}^{\rm o}_0$ at $t=0$ and $\bf{x}^{\rm o}_{1}$ and $t=t_1$.
Along this partition, multiple forward trajectories will be sampled recursively, by drawing from $p(\bf{x}_{k} \given \bf{x}_{{k - 1}}, \bs{\theta})$, and weighting according to 
\begin{equation}
    W_k(\bf{x}_{{k - 1}}, \bf{x}_{k}) =  \frac{q(\bf{x}^{\rm o}_{1} \given \bf{x}_{{k}})}{q(\bf{x}^{\rm o}_{1} \given \bf{x}_{{k - 1}})}, \quad \text{for } k = 1, \ldots, A, \label{eq:seq-weights}
\end{equation}
where $q(\bf{x}^{\rm o}_{1} \given \bf{x}_{{k}})$, for $k = 0, \ldots, A$ are positive functions. This forms the basis of the BPF \citep{del2015sequential}, albeit for the case of exact observations (i.e. when  $\bf{x}^{\rm o}_{1}$ is observed without noise). Note that the weighting function $q(\bf{x}^{\rm o}_{1} \given \bf{x}_{{A}})$ at time $\tau_A$ can be omitted because $\bf{x}_{A}$ is fixed to be equal to $\bf{x}^{\rm o}_{1}$. As alluded to previously, instead of fixing the last state as in the BPF, we perform one more forward propagation step, and hence include $q(\bf{x}^{\rm o}_{1} \given \bf{x}_{{A}})$. A suitable choice for this weighting function is important because it greatly affects the efficiency of the inference. We discuss possible options later in this section.
Through the use of the lookahead mechanism, the intermediate samples between two observations, $\bf{x}_{1}, \ldots, \bf{x}_{{A - 1}}$ (excluding the final sample $\bf{x}_{A}$) are given weights according to how consistent they are with the subsequent observation $\bf{x}^{\rm o}_{1}$. However, for the final sample $\bf{x}_{{A}}$, this approach is invalid because the transition density at $\tau_{A} \equiv t_1$ is a Dirac mass at the observation point. To this end we propose to take $q(\bf{x}^{\rm o}_{1} \given \bf{x}_{A})$ to be equal to $q(\bf{x}^{\rm o}_{1} \given \bf{x}_{{A - 1}})$, which is a plausible choice when the integration timestep $\tau_{A} - \tau_{A-1} = h$ is small. This way the particles $(\bf{x}^{1:P}_{{A}})$ that are close to the observational point $\bf{x}^{\rm o}_{1}$ will be given greater weights, as opposed to those that are further away. Another possible choice for $q(\bf{x}^{\rm o}_{1} \given \bf{x}_{A})$ may be a decaying function of the distance between $\bf{x}^{\rm o}_{1}$ and $\bf{x}_{{A}}$, for example the Gaussian kernel, or the inverse of the (squared) Euclidean distance.
The conditional distribution of the intermediate points, given the subsequent observation, admits the following form
\begin{align}
    p(\bf{x}_{1 : A} \given \bf{x}^{\rm o}_0, \bf{x}^{\rm o}_{1}) & \propto q(\bf{x}^{\rm o}_{1} \given \bf{x}_{A}) p(\bf{x}_{1 : A} \given \bf{x}^{\rm o}_0) \frac{q(\bf{x}^{\rm o}_{1} \given \bf{x}^{\rm o}_0)}{q(\bf{x}^{\rm o}_{1} \given \bf{x}_{A})} \prod_{k = 1}^A W_k(\bf{x}_{{k-1}}, \bf{x}_{k}) \nonumber \\
    & \propto \prod_{k = 1}^A W_k(\bf{x}_{{k-1}}, \bf{x}_{k}) p(\bf{x}_{{k}} \given \bf{x}_{{k-1}}). \label{eq:sis} 
\end{align}
Note that 
\begin{equation}
    \frac{q(\bf{x}^{\rm o}_{1} \given \bf{x}^{\rm o}_0)}{q(\bf{x}^{\rm o}_{1} \given \bf{x}_{t_1})} \prod_{k = 1}^A W_k(\bf{x}_{{k-1}}, \bf{x}_{k}) = 1. \nonumber
\end{equation}
For a generic observation interval $[t_i, t_{i + 1}]$, the conditional distribution of the intermediate points $\bf{x}_{{iA + 1}}, \ldots, \bf{x}_{{(i + 1)A}}$ is similar to \eqref{eq:sis}, except that the state $\bf{x}_{{iA}}$ at time $\tau_{iA}$ is not equal to the observation $\bf{x}^{\rm o}_{i}$, but to the final value that was simulated on the previous interval $[t_{i - 1}, t_{i}]$. This is different from the BPF case where the particles are initialized to start from the observation at time $t_i$. Having derived the conditional distribution of the intermediate points on the first interval, we can easily extend \eqref{eq:sis} to the complete discretization $(\tau_0, \tau_1, \ldots, \tau_{A}, \ldots \tau_{N})$ as follows
\begin{align}
    p(\bf{x}_{{1} : {N}} \given \mathbf{x}^{\rm o}) & = \prod_{i = 0}^{n - 1} p(\bf{x}_{{iA + 1}: {(i + 1) A}} \given \bf{x}_{{iA}}, \bf{x}^{\rm o}_{{i + 1}}) \nonumber \\ & \propto \prod_{i = 0}^{n - 1} \prod_{k = iA + 1}^{(i + 1)A}  W_k(\bf{x}_{{k-1}}, \bf{x}_{k}) p(\bf{x}_{{k}} \given \bf{x}_{{k-1}}). \label{eq:forwardcomplete}
\end{align}
Equation \eqref{eq:forwardcomplete} readily implies a sequential scheme where one can incrementally sample $\bf{x}_{{k}} \sim p(\bf{x}_{{k}} \given \bf{x}_{{k-1}})$ and weight that sample by $W_k(\bf{x}_{{k - 1}}, \bf{x}_{k})$, for $k = 1, \ldots, N$. Assume that we have simulated a set of $P$ particles up to time $t_n$, then we have the particle system $(\bf{x}^{1:P}_{1:N}, \omega^{1:P}_{1:N})$ approximating the lookahead densities
\begin{equation}
    \hat{p}({\rm d}\bf{x}_{{iA + k}} \given \bf{x}^{\rm o}_{0:t_{i + 1}}) = \sum_{j = 1}^P \omega_{iA + k}^j \delta_{\bf{x}^j_{{iA + k}}} ({\rm d}\bf{x}_{{iA + k}}) \quad \text{for } k = 1, \ldots, A, \label{eq:part-approx-lookahead}
\end{equation}
for $i = 0 \ldots, n - 1$, where $\omega_k^j$ denote the normalized particle weights. Due to the way the weights $W_k(\bf{x}_{{k - 1}}, \bf{x}_{k})$ are defined in \eqref{eq:seq-weights}, it is worth noting that in the sequential-importance-sampling scheme, the weight of the sample $\bf{x}_{{iA + k}} \sim p(\bf{x}_{{iA + k}} \given \bf{x}_{{iA + k-1}})$ is proportional to $q(\bf{x}^{\rm o}_{{i + 1}} \given \bf{x}_{{iA + k}})$ (see also Algorithm 3 in \citealp{del2015sequential} and the subsequent note on extending it to a time series of observations). In Algorithm \ref{algo:SIS} we present the lookahead SIS algorithm corresponding to equation \eqref{eq:forwardcomplete} for a set of $P$ particles, an illustration of which can be seen in Figure \ref{fig:lookahead-sis}.

Having approximated the lookahead densities up to time $t_n$, we can now obtain an approximation of the so called backward kernel, to evolve backward in time and obtain a single trajectory that we use in ABC-SMC to decide on whether to accept or reject the associated parameter. Backward simulation methods for Monte Carlo \citep{lindsten2013backward}, as well as the derivation of the backward simulation particle smoother for the finer discretization of $[0, t_n]$, is the topic of the following section.
\begin{algorithm}[ht]

    \caption{Lookahead SIS $(\bf{x}^{\rm o}, \bs{\theta})$}\label{algo:SIS}
    \begin{algorithmic}[1]
        \FOR{$j = 1$ to $P$ in parallel}
        \FOR{$i = 0$ to $n - 1$} 
            \FOR{$k = 1$ to $A$}
                \STATE Sample $\bf{x}_{{iA + k}}^j \sim p( \bf{x}_{{iA + k}} \given \bf{x}_{{iA + k - 1}}^j, \boldsymbol{\theta})$ with timestep $h$. 
                \STATE Append the sample to the state history $\bf{x}_{{0}:{iA + k}}^j = \{\bf{x}_{{0}:{iA + k - 1}}^j, \bf{x}_{{iA + k}}^j\}$.
                \STATE Calculate the weight of the particle according to 
                $\omega^j_{iA + k} \propto q(\bf{x}^{\rm o}_{{i + 1}} \given \bf{x}^j_{{iA + k}})$. 
                $\mathcal{N}(\bf{x}^{\rm o}_{{i + 1}} \given \bf{x}^j_{{iA + k}} + \bs{\mu}(\bf{x}^j_{{iA + k}}, \bs{\theta})(t_{i + 1} - \tau_{iA + k}), \bs{\sigma}(\bf{x}^j_{{iA + k}}, \bs{\theta}) \bs{\sigma}^T(\bf{x}^j_{{iA + k}}, \bs{\theta})(t_{i + 1} - \tau_{iA + k}))$. 
            \ENDFOR
            \ENDFOR
        \ENDFOR
    \STATE \textbf{Output:} Particle system $(\bf{x}^{1:P}_{1:N}, \omega^{1:P}_{1:N})$
    \end{algorithmic}
\end{algorithm}
\subsection{Extracting a single trajectory using particle smoothing}
\begin{figure}
    \centering
    \includegraphics[scale=0.65]{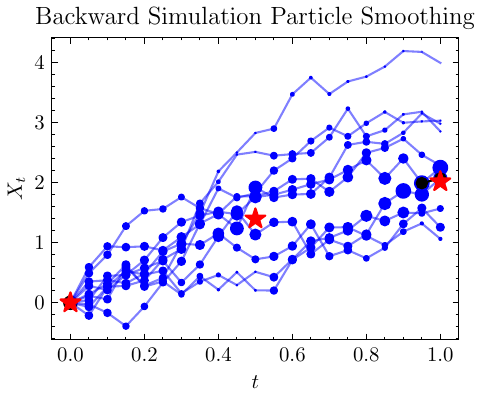}
    \includegraphics[scale=0.65]{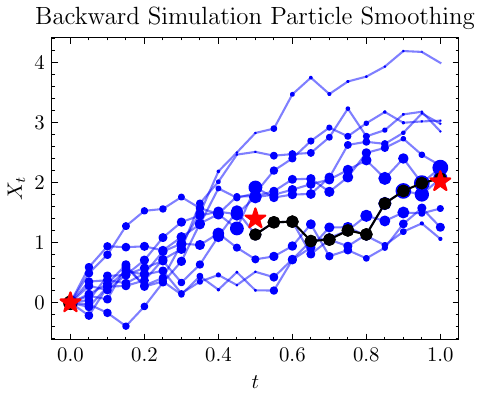}
    \includegraphics[scale=0.65]{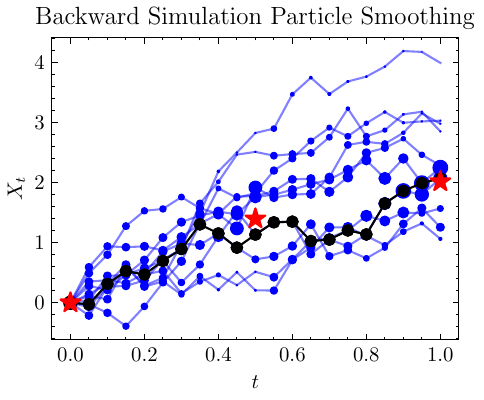}
    \caption{Evolution of a trajectory that is obtained via the backward simulation particle smoother on the particle system in Figure \ref{fig:lookahead-sis}. Left: a particle  is chosen at random according to equation \eqref{eq:end-backward} and a single step backward is taken according to equation \eqref{eq:recursion-backwards} (the chosen particle is in black). Middle: the backward recursion is taken up to time $t_1$ (in black), and it can be seen that the backward trajectory is passing close to the observation. The right panel depicts the complete backward trajectory in black.}
    \label{fig:bsps}
\end{figure}

This section describes how to make use of the particles generated in the forward direction (see Algorithm \ref{algo:SIS}) to select a suitable simulated trajectory for use within ABC-SMC. Backward simulation (BS) is a technique that is used within SMC to address the smoothing problem for models that have latent stochastic processes, i.e. state-space models \citep{lindsten2013backward, sarkka2013bayesian}. In a state-space model, it is assumed that a dynamical system evolves according to a Markovian (latent) stochastic process having state $\bf{x}_t$ at time $t$ which is not directly measured; rather, $\bf{y}_t$ is observed as some function of the latent $\bf{x}_t$. Smoothing refers to the problem of estimating the distribution of the latent state at a particular time, given all of the observations up to some later time. More formally, given a set of observations $\bf{y} = (\bf{y}_{t_1}, \ldots, \bf{y}_{t_n}) $, smoothing addresses the estimation of the marginal distributions $p(\bf{x}_{t_k} \given \bf{y})$ for $k = 1, \ldots, n$, or sampling from the entire smoothing density $p(\bf{x}_{t_1}, \ldots, \bf{x}_{t_n} \given \bf{y})$. Backward simulation is based on the forward-backward idea, and assumes that Bayesian filtering has already been performed on the entire collection of observations, leading to an approximate representation of the densities $p(\bf{x}_t \given \bf{y}_{1:t})$ for each time step $t \in (t_1, \ldots, t_n )$, consisting of weighted particles $(\bf{x}^{1:P}_{t}, \omega^{1:P}_t)$. The primary goal of BS is to obtain sample realizations from the smoothing density to gain insight about the latent stochastic process. The smoothing density can be factorized as
\begin{equation}
    p(\bf{x}_{t_1 : t_n} \given \bf{y}) = p(\bf{x}_{t_n} \given \bf{y}) \prod_{i = 1}^n p(\bf{x}_{t_i} \given \bf{x}_{t_{i + 1}:t_{n}}, \bf{y})  \label{eq:jsd}
\end{equation}
where, under the Markovian assumption of the model, one can write  
\begin{align}
    p(\bf{x}_{t_i} \given \bf{x}_{t_{i + 1} : t_n}, \bf{y}) & = p(\bf{x}_{t_i} \given \bf{x}_{t_{i + 1}}, \bf{y}_{t_1 : t_i}), \nonumber \\
    & \propto p(\bf{x}_{t_i} \given \bf{y}_{t_1 : t_i}) p(\bf{x}_{t_{i + 1}} \given \bf{x}_{t_{i}}). \label{eq:backward-eq}
\end{align}
This equation additionally shows that, conditional on $\bf{y}$, $(\bf{x}_{t_{1}}, \ldots, \bf{x}_{t_{n}})$ is an inhomogeneous Markov process. 
In our case we do not have a latent process as in the state-space model, but rather exact {(without measurement error)}, discrete-time observations of an SDE. Still, we can utilize backward simulation because of the way that we have constructed the lookahead sequential importance sampling (Lookahead SIS, Algorithm \ref{algo:SIS}) scheme. Moreover, the output of backward simulation will be the simulated trajectories that will be used in the ABC algorithm. Next, we focus on the backward simulation step where we derive the backward recursion along the finer discretization that will be used to generate smoothed trajectories. The key ingredient in this second step is the backward kernel
\begin{equation}
    \mathcal{B}_{iA + k}({\rm d}\bf{x} \given \bf{x}_{{iA + k + 1}}) = \mathbb{P}(\bf{x}_{{iA + k}} \in {\rm d}\bf{x} \given \bf{x}_{{iA + k + 1}}, \bf{x}^{\rm o}_{0:t_{i + 1}}), \label{eq:backward-kernel}
\end{equation}
which admits the following density
\begin{equation}
    p(\bf{x}_{{iA + k}} \given \bf{x}_{{iA + k + 1}}, \bf{x}^{\rm o}_{0:t_{i + 1}}) \propto p(\bf{x}_{{iA + k + 1}} \given \bf{x}_{{iA + k}}) p(\bf{x}_{{iA + k}} \given \bf{x}^{\rm o}_{0:t_{i + 1}}). \label{eq:backward-density}
\end{equation}
Using the backward kernel we can get the following expression for the backward recursion
\begin{align}
    p(\bf{x}_{{iA + k}:{N}} \given \bf{x}^{\rm o}_{0:t_n}) & = p(\bf{x}_{{iA + k}} \given \bf{x}_{{iA + k + 1}}, \bf{x}^{\rm o}_{0:t_{i + 1}}) p(\bf{x}_{{iA + k + 1} : {N}} \given \bf{x}^{\rm o}_{0:t_{n}}), \label{eq:backward-recursion}
\end{align}
starting from the lookahead density $p(\bf{x}_{{N}} \given \bf{x}^{\rm o}_{0:t_{n}})$ at time $t_n$. Equation \eqref{eq:backward-recursion} implies the following scheme: (i) sample 
\begin{equation}
\tilde{\bf{x}}_{{N}} \sim p(\bf{x}_{{N}} \given \bf{x}^{\rm o}_{0:t_n}),   \label{eq:end-backward}
\end{equation}
and then going backwards in time, (ii) incrementally sample 
\begin{equation}
\tilde{\bf{x}}_{{iA + k}} \sim p(\bf{x}_{{iA + k}} \given \bf{x}_{{iA + k + 1}}, \bf{x}^{\rm o}_{0:t_{i + 1}}).  \label{eq:recursion-backwards}
\end{equation}
After a complete backwards sweep, the (backward) trajectory $(\tilde{\bf{x}}_{0}, \ldots, \tilde{\bf{x}}_{{N}})$ is by construction a realization from the smoothing density $p( \bf{x} \given \bf{x}^{\rm o})$. 
An important property of the backward kernel density is that at time $\tau_{iA + k}$, it depends only on the transition density $p(\bf{x}_{{iA + k + 1}} \given \bf{x}_{{iA + k}})$, which we are able to approximate for a small timestep $h$, and on $p(\bf{x}_{{iA + k}} \given \bf{x}^{\rm o}_{0:t_{i + 1}})$, which is approximated by the weighted particle system obtained in the forward direction. To this end, to utilize the backward recursion, the lookahead densities $p(\bf{x}_{{iA + k}} \given \bf{x}^{\rm o}_{0:t_{i + 1}})$ must first be computed for $i = 0 \ldots, n - 1$ and $k = 1, \ldots, A$. By substituting the particle approximation of the lookahead densities \eqref{eq:part-approx-lookahead} into the backward density \eqref{eq:backward-density} we have the following approximation to the backward kernel
\begin{equation}
    \hat{\mathcal{B}}_{iA + k}({\rm d}\bf{x}_{{iA + k}} \given \bf{x}_{{iA + k + 1}}) = \sum_{j = 1}^P \frac{\omega_{iA + k}^j p(\bf{x}_{{iA + k + 1}} \given \bf{x}^j_{{iA + k}})}{\sum_{l = 1}^P \omega_{iA + k}^l p(\bf{x}_{{iA + k + 1}} \given \bf{x}^l_{{iA + k}})} \delta_{\bf{x}^j_{{iA + k}}} ({\rm d}\bf{x}_{{iA + k}}). \label{eq:app-kernel}
\end{equation}
Algorithm \ref{algo:BSPS} presents the complete backward recursion given a weighted particle system obtained from the lookahead SIS algorithm. A visual illustration is given in Figure \ref{fig:bsps}. Notice that while the complete backward recursion results in a trajectory $\tilde{\bf{x}}$ of length $nA + 1$, we only need the values at the observational timepoints $(t_1, ..., t_n)$. As such, we downsample $\tilde{\bf{x}}$ at these points. Inspired by this, we explored taking larger time steps backward, despite deriving our recursions, \eqref{eq:end-backward}-\eqref{eq:recursion-backwards}, on a fine grid. For instance, rather than retracing steps on the fine grid at intervals of $\Delta t / A$ (which involves $A$ steps), we considered intervals of larger lengths, for example $\Delta t / 2$, transitioning from $t_i$ to $t_{i - A / 2}$, and then to $t_{i - 1}$ (a total of 2 steps). We experimented with different step sizes and found that stepping backward directly from $t_i$ to $t_{i - 1}$ yielded the most consistent trajectories. As a result, our simulation studies adopt this method. Yet, the forward trajectories maintain their precision, given the forward simulator integrates at $h = \Delta t / A$.

\begin{algorithm}[tp]

    \caption{BSPS: backward-simulation particle smoother $((\bf{x}^{1:P}_{1:N}, \omega^{1:P}_{1:N}), \boldsymbol{\theta})$}\label{algo:BSPS}
    \begin{algorithmic}[1]
    \STATE Sample particle index $j \sim \mathcal{M}(\{\omega_N^i\}_{i=1}^P)$ and set $\tilde{\bf{x}}_{N} = \bf{x}^j_{N}$.
        \FOR{$k = N - 1$ to $1$}
            \FOR{$j = 1$ to $P$}
                \STATE Compute $\tilde{\omega}_k^j \propto \omega_k^j p(\tilde{\bf{x}}_{{k + 1}} \given \bf{x}_{{k}}^j, \bs{\theta})$. 
            \ENDFOR
            \STATE Normalize the smoothing weights $\{\tilde{\omega}_k^{1:P}\}$ to sum to unity.
            \STATE Sample particle index $j \sim \mathcal{M}(\{\tilde{\omega}_k^{1:P}\})$ and set $\tilde{\bf{x}}_{{k}} = \bf{x}_{{k}}^j$.
            \STATE Append the sample to the state history $\tilde{\bf{x}}_{{k}: {N}} = \{\tilde{\bf{x}}_{{k}}, \tilde{\bf{x}}_{{k + 1}: {N}}\}$.
        \ENDFOR
        \STATE \textbf{Output:} Backward trajectory $\tilde{\bf{x}}_{{0}: {N}}$.
    \end{algorithmic}
\end{algorithm}

In summary, given a parameter proposal $\bs{\theta}^*$ and the observation $\bf{x}^{\rm o}$, the simulation of a backward trajectory $\tilde{\bf{x}}$ is achieved by executing Algorithms \ref{algo:SIS} and \ref{algo:BSPS}. Algorithm \ref{algo:SIS} returns the particle system that approximates the lookahead densities, and this same particle system is used as input to Algorithm \ref{algo:BSPS} to approximate the backward kernel density. We denote by $\tilde{\bf{x}} \sim p(\bf{x} \given \bs{\theta}^*, \bf{x}^{\rm o})$ the process of sampling a backward trajectory by executing the two algorithms. This $\tilde{\bf{x}}$ is the trajectory for which the ABC summary statistic $s_t^i=S(\tilde{\bf{x}})$ is calculated and evaluated in line 18 of the ABC-SMC Algorithm \ref{algo:sl-abs-smc}. In the next two sections we clarify further steps that are necessary for the adaptation of ABC-SMC to our specific context. In particular, section \ref{sec:synthetic-like} discusses how to appropriately weight the accepted parameter particles in ABC-SMC. Section \ref{sec:summaries} discusses the construction of ABC summary statistics that are suitable for SDE inference. 

\section{Intractable importance ratio and its approximation}\label{sec:synthetic-like}

Throughout section \eqref{sec:backward-sims} we have addressed the problem of simulating paths $\tilde{\bf{x}}$ using a data-conditional forward-backward procedure. It now remains to be seen how to (i) construct suitable summary statistics that we apply to both $\bf{\tilde{x}}$ and data $\bf{x}^o$, and (ii) how to correct for the fact that $\bf{\tilde{x}}$ is not simulated from the forward model. We now consider (ii), and address (i) in section \ref{sec:summaries}.
With a standard, forward only, simulator we would only need to care for $p(\mathbf{s} \given \bs{\theta})$, with $\bf{s}=S(\bf{\tilde{x}})$ and $S(\cdot)$ a  function still to be determined (to be addressed in section \ref{sec:summaries}). 
The utilization of the backward simulator effectively changes the distribution of the simulated trajectories. The summary of this backward trajectory is no longer distributed according to $p(\mathbf{s} \given \bs{\theta})$, but rather according to another density that is conditional on observed data and which we denote by $p(\mathbf{s} \given \bs{\theta}, \bf{x}^{\rm o})$ (this density is implicitly defined as that of the summary statistic of the backward trajectory sampled from the joint smoothing density $p(\mathbf{x} \given \bs{\theta}, \bf{x}^{\rm o})$). Hence, the importance weight of the parameter-summary pair takes a different form and includes the intractable summary likelihoods of both the forward and the backward simulator. More precisely, the sampling density $g(\mathbf{s}, \bs{\theta})$ is no longer $p(\mathbf{s} \given \bs{\theta})g(\bs{\theta})$, but rather $p(\mathbf{s} \given \bs{\theta}, \mathbf{x}^{\rm o})g(\bs{\theta})$. We denote this new sampling density by $g(\mathbf{s}, \bs{\theta} \given \mathbf{x}^{\rm o})$ to emphasize the dependence on the observed data. For a proposed parameter-summary pair the importance weight becomes 
\begin{equation}
    \frac{\pi_\epsilon(\bs{\theta}^*, \mathbf{s}^* \given \mathbf{s}^{\rm o})}{g(\bs{\theta}^*, \mathbf{s}^* \given \mathbf{x}^{\rm o})} \propto \frac{ \mathbbm{1}( \Vert \mathbf{s} - \mathbf{s}^{\rm o} \Vert \leq \epsilon)  \pi(\bs{\theta}^*)}{g(\bs{\theta}^*)} \frac{p(\mathbf{s}^* \given \bs{\theta}^*)}{p(\mathbf{s}^* \given \bs{\theta}^*, \mathbf{x}^{\rm o})}. \label{eq:accept-probability-looakhead}
\end{equation}
Unlike the corresponding ratio for forward simulators in \eqref{eq:complete-acc-prob}, the importance weight in \eqref{eq:accept-probability-looakhead} is intractable. 
Below, we focus on developing a computationally efficient procedure for approximating the ratio \eqref{eq:accept-probability-looakhead}, by making use of the Synthetic Likelihood (SL) method. The SL method was first proposed in \cite{wood2010statistical} to perform inference for parameters of
computer-based models with an intractable likelihood. SL is characterized by the assumption that the summary statistic $\bf{s}$ follows a multivariate normal distribution with unknown mean $\bs{\mu}_{\bs{\theta}}$ and covariance $\bs{\Sigma}_{\bs{\theta}}$. These can in turn be estimated by implicitly sampling a set of $P$ summary statistics for a particular parameter $\bs{\theta}$, $\bf{s}^j \sim p(\mathbf{s} \given \bs{\theta})$ for $j = 1, \ldots, P$, and then computing their empirical mean $\bs{\mu}_{P, \bs{\theta}}$ and covariance $\bs{\Sigma}_{P, \bs{\theta}}$. This results in the approximation $p(\bf{s} \given \bs{\theta}) \approx \mathcal{N}(\bf{s} \given \bs{\mu}_{P, \bs{\theta}}, \bs{\Sigma}_{P, \bs{\theta}})$. To construct an appropriate SL approximation of $p(\bf{s} \given \boldsymbol{\theta}, \bf{x}^{\rm o})$, we need to look more closely into the particle approximation to the joint smoothing density of the backward trajectory. By plugging the approximations of the lookahead densities \eqref{eq:part-approx-lookahead} and the backward kernels \eqref{eq:backward-kernel} into the joint smoothing density \eqref{eq:jsd} along the fine discretization, we get the following approximation 
    \begin{equation}
    {\hat{p}({\rm d} \bf{x} \given \bf{x}^{\rm o}, \bs{\theta}) = \sum_{j_1 = 1}^P \cdots \sum_{j_n = 1}^P \left( \prod_{i = 1}^{n - 1} \frac{\omega_{iA}^{j_i} p(\bf{x}_{{(i+1)A}}^{j_{i + 1}} \given \bf{x}_{iA}^{j_i}, \bs{\theta}) }{\sum_{l=1}^P \omega_{iA}^{l} p(\bf{x}_{{(i+1)A}}^{j_{i + 1}} \given \bf{x}_{{iA}}^{l}, \bs{\theta})} \right) \omega_N^{j_n} \delta_{\mathbf{x}_{A}^{j_1}, \ldots, \mathbf{x}_{N}^{j_n}}({\rm d} \mathbf{x}).}  \label{eq:approx-jsd}
\end{equation}
This equation defines a discrete distribution on {$\mathcal{X}^n$}
, and can be understood as follows: for each observational time {${t_i}$}, $i = 1, \ldots, n$ along the coarse discretization 
, the particles {$(\mathbf{x}_{iA}^{1:P})$} 
generated by Lookahead SIS is a set in $\mathcal{X}$ of cardinality $P$. By picking one particle at each time point, we obtain a particle trajectory, i.e. a point {$(\mathbf{x}_{A}^{j_1}, \mathbf{x}_{2A}^{j_2}, \ldots, \mathbf{x}_{N}^{j_n}) \in \mathcal{X}^n$} 
with entries obtained at the $n$ observational time points. We get $P^n$ such trajectories by letting the indices $j_1, \ldots, j_n$ 
range from $1$ to $P$. Clearly the form of \eqref{eq:approx-jsd} is computationally intensive to evaluate, however, it provides a connection to the backward simulator (Algorithm \ref{algo:BSPS}). Implicitly, the particle approximation to the joint smoothing density \eqref{eq:approx-jsd} depends on the particle system from the forward direction, i.e. $\hat{p}({\rm d} \bf{x} \given \bf{x}^{\rm o}, \bs{\theta}) = \hat{p}({\rm d} \bf{x} \given \bf{x}^{\rm o}, \bs{\theta}, (\bf{x}^{1:P}_{1:N}, \omega^{1:P}_{1:N}))$. Therefore, conditionally on the particle system from Lookahead SIS, the backward simulator generates i.i.d. Markovian samples from the distribution \eqref{eq:approx-jsd}. Thus, to appropriately approximate $p(\bf{s} \given \boldsymbol{\theta}, \bf{x}^{\rm o})$ by a Gaussian likelihood, the weighted particle system that is obtained from Lookahead SIS needs to remain fixed whilst the backward trajectories are being sampled. The larger the number of samples $P$, the better the approximation to the summary likelihood, albeit at a higher simulation cost. In our applications, we set the number of simulations for SL to be the same as the number of particles for the lookahead simulator, which we found to work well in the empirical examples. Hence if $P$ is set to a high number, the cost per one forward-backward simulation will be increased.  Similarly to the forward summary likelihood, the backward summary likelihood can be approximated by sampling a set of {$P$ data-conditional} summary statistics $\tilde{\bf{s}}^j \sim p(\mathbf{s} \given \bs{\theta}, \mathbf{x}^{\rm o})$ and then computing their empirical mean $\tilde{\bs{\mu}}_{P, \bs{\theta}}$ and covariance $\tilde{\bs{\Sigma}}_{P, \bs{\theta}}$. The resulting approximation is $p(\bf{s} \given \bs{\theta}, \mathbf{x}^{\rm o}) \approx \mathcal{N}(\bf{s} \given \tilde{\bs{\mu}}_{P, \bs{\theta}}, \tilde{\bs{\Sigma}}_{P, \bs{\theta}})$. The importance weight for a parameter-summary pair $(\bs{\theta}, \mathbf{s})$ can now be approximated as
\begin{equation}
    w(\bs{\theta}, \mathbf{s}) \appropto \frac{\mathbbm{1}( \Vert \mathbf{s} - \mathbf{s}^{\rm o} \Vert \leq \epsilon)  \pi(\bs{\theta})}{  g(\bs{\theta})} \frac{\mathcal{N}(\bf{s} \given \bs{\mu}_{P, \bs{\theta}}, \bs{\Sigma}_{P, \bs{\theta}})}{\mathcal{N}(\bf{s} \given \tilde{\bs{\mu}}_{P, \bs{\theta}}, \tilde{\bs{\Sigma}}_{P, \bs{\theta}})}. \label{eq:is-approx}
\end{equation} 
Crucially, the form of the importance weight in \eqref{eq:is-approx} indicates that the ratio of the intractable likelihoods need only be computed when the simulated summary is close enough to the observed summary, i.e. $\mathbbm{1}( \Vert \mathbf{s} - \mathbf{s}^{\rm o} \Vert \leq \epsilon) = 1$, otherwise the importance weight can be immediately set to zero without the need to compute the synthetic likelihoods. The procedure for computing the ratio is outlined in Algorithm \ref{algo:sl-approx}. 
\begin{algorithm}

    \caption{Synthetic likelihood approximation $((\bf{x}^{1:P}_{1:N}, \omega^{1:P}_{1:N}), \boldsymbol{\theta}, \mathbf{s})$ }\label{algo:sl-approx}
    \begin{algorithmic}[1]
            \FOR{$j = 1$ to $P$}
                \STATE Summarize the genealogy of the $j$\textsuperscript{th} particle $\mathbf{s}^{j} = S(\mathbf{x}^{j})$.
                \STATE Sample $\tilde{\mathbf{x}}^{j} \sim \hat{p}(\bf{x} \given \bf{x}^{\rm o}, \bs{\theta}, (\bf{x}^{1:P}_{1:N}, \omega^{1:P}_{1:N}))$ and summarize $\tilde{\mathbf{s}}^{j} = S(\tilde{\mathbf{x}}^{j})$.
            \ENDFOR
            \STATE Calculate the empirical means and covariance matrices from $\mathbf{s}^{1:P}$ and $\tilde{\mathbf{s}}^{1:P}$, respectively: $\bs{\mu}_{P, \bs{\theta}}$ (resp. $\tilde{\bs{\mu}}_{P, \bs{\theta}}$) and  $\bs{\Sigma}_{P, \bs{\theta}}$ (resp. $\tilde{\bs{\Sigma}}_{P, \bs{\theta}}$).
            \STATE \textbf{Output:} Approximate ratio $\mathcal{N}(\mathbf{s} \given \bs{\mu}_{P, \bs{\theta}}, \bs{\Sigma}_{P, \bs{\theta}}) / \mathcal{N}(\mathbf{s} \given \tilde{\bs{\mu}}_{P, \bs{\theta}}, \tilde{\bs{\Sigma}}_{P, \bs{\theta}})$.
    \end{algorithmic}
\end{algorithm}
In the supplementary material, we explore a crucial property of the approximate importance weight, as defined in \eqref{eq:is-approx}, which allows for the exclusion of implausible parameters, independent of the ABC threshold \(\epsilon\) value. This method hinges on the variance of the weights of the particle system, obtained from Lookahead SIS, and its impact on the covariance matrix \(\tilde{\bs{\Sigma}}_{P, \bs{\theta}}\) of the synthetic likelihood found in the denominator of \eqref{eq:is-approx}.

\section{Automatic summary statistics for SDEs}
\label{sec:summaries}

In this section we address the construction of the summary function $S(\cdot)$, that has so far been left unspecified.
One of the main challenges when employing ABC algorithms in practice is the choice of summary statistics that retain enough information about $\bm{\theta}$. In fact, high-dimensional data, like the numerical solutions to SDEs, can require a large threshold $\epsilon$, implying an inflated posterior approximation, unless data dimension is first reduced via summary statistics. On the other hand, a too small $\epsilon$ will result in very low acceptance rates. To deal with this problem, the typical approach is to reduce the data to lower dimensional summary statistics and instead compare the simulated summaries $S(\bf{x})$ to the observed summaries $S(\bf{x}^{\rm o})$, though this is not the only possibility, see \cite{drovandi2022comparison} for an extensive comparison on alternatives. 
A popular approach for the selection of summary statistics is to use regression models. The idea  was first introduced in \cite{fearnhead2012constructing}, where it was shown that the optimal summary statistic under the quadratic loss is the posterior mean. It was proven that when choosing $S(\bf{x}) = \mathbb{E}(\bs{\theta} \given \bf{x})$, the ABC posterior has the same mean as the exact posterior in the limit $\epsilon \to 0$. Given this result, the authors propose to estimate the posterior mean by fitting a linear regression model on a batch of prior-predictive simulations. For the $j$th parameter ($j = 1, \ldots, p$) the following linear regression model is built $\theta^i_j = \mathbb{E}(\theta_j \given \bf{x}^i) + \xi^i_j = b_{0_j} + b_j h(\bf{x}^i) + \xi^i_j, j = 1, \ldots, p$, with $\xi^i_j$ some mean-zero random term with constant variance and $(\bs{\theta}^i, \bf{x}^i)_i$ are samples from the prior-predictive distribution. The $p$ different regression models are estimated separately by least squares, and the corresponding fit $\hat{b}_{0_j} + \hat{b}_j h(\bf{x})$ provides an estimate for $\mathbb{E}(\theta_j \given \bf{x})$ which can then be used as a summary statistic because of the optimality of the posterior mean. These least squares fits are performed before the start of the ABC algorithm and hence the summary statistics function remains fixed throughout. This work was then further developed and fully automatised by \cite{jiang2017learning}, 
who model the regression function via a deep neural network $\bs{\theta}^i = \mathbb{E}(\bs{\theta} \given \bf{x}^i) + \xi^i = f_\beta(\bf{x}^i) + \xi^i$, for the complete multidimensional parameter $\bs{\theta}^i$, and where $f_\beta(\cdot)$ is a neural network parametrized by weights $\beta$. Due to the greater representational power of neural networks, the deep learning approach outperforms the linear regression approach, although at the expense of higher computational cost. 
Partially exchangeable networks (PEN, \citealp{wiqvist2019partially}) are a member of the family of invariant neural networks. They were introduced to ease the process of ABC inference by automatically learning the posterior mean as a summary statistic, by leveraging the partially exchangeable structure of Markov chains. Therefore, for SDE inference, PENs offer a powerful way to learn ABC summary statistics, at a much lower computational cost (i.e. smaller training data) than in \cite{jiang2017learning}, since PENs do not have to ``learn'' the Markovian structure of the data, given that the network itself is designed to accommodate such a framework. \cite{wiqvist2019partially} show that for models which are invariant under the action of a subset of the symmetric group, called $d$-block-switch transformations, the neural network should also be $d$-block-switch invariant. To this end, they propose the following regression model for the posterior mean
\begin{equation}
    \bs{\theta}^i = \mathbb{E}(\bs{\theta} \given \bf{x}^i) + \xi^i = \rho_{{\beta}_\rho}\left(\bf{x}^i_{1:d}, \sum_{l = 1}^{n - d} \phi_{{\beta}_\phi}(\bf{x}^i_{l:l+d})\right) + \xi^i. \label{eq:posterior-mean-learning}
\end{equation}
PEN is a composition of two neural networks:
$\phi(\cdot)$ is called  the inner network, which maps a subsequence $\bf{x}_{i:i+d}$ into some representation $\phi(\bf{x}_{i:i+d})$ and $\rho$ is the outer network that maps the first $d$ symbols of the input and the sum of all $d$-length subsequences to the output. Here, ${\beta}_\phi$ are the weights for the inner network, and ${\beta}_\rho$ are the weights for the outer network that maps its arguments into the posterior mean of the unknown parameters. 
The summary statistics estimator can be trained before running the inference algorithm. However, for this strategy to be effective, very many samples from the prior-predictive distribution are required, and this number is affected by how ``vague'' the prior is. This is the approach taken in \cite{picchini2014inference, wiqvist2019partially, fearnhead2012constructing, jiang2017learning, chan2018likelihood}. Recently, \cite{chen2020neural} proposed a dynamic learning strategy, where the main idea is to learn the summary statistics and the posterior density at the same time, over multiple rounds, which we also advocate here. More precisely, at round $t$, the current summary statistics network $S_t(\cdot)$ is used to approximate the ABC posterior density, and then the summary statistics network is retrained on all the data obtained up to round $t$, and the newly trained network $S_{t + 1}(\cdot)$ is used in round $t + 1$. 

{In order to learn the summary statistics in our approach, we denote with $\mathcal{D}$ the set containing training data.
In the standard (forward) ABC-SMC approach, the dataset $\mathcal{D}$ is progressively populated with accepted parameters and their corresponding simulated trajectories. In our approach, however, this step warrants careful attention. For a given parameter, a set of $P$ forward trajectories are sampled with Algorithm \ref{algo:SIS} and one backward trajectory is sampled with Algorithm \ref{algo:BSPS}. The backward trajectory is summarized and then passed to the ABC accept/reject step, but it is not this trajectory that is stored into $\mathcal{D}$. The distribution of the backward trajectories is markedly different from that of the forward trajectories, particularly for parameters that do not come from the posterior (see Section 1 of the Supplementary Material for a detailed discussion). Therefore, to keep the learning of the summary statistics consistent, we select one of the $P$ \textit{forward} trajectories according to how close it is to the observed trajectory. More precisely, we downsample the $P$ forward trajectories and find the one with the minimum Euclidean distance to $\bf{x}^{\rm o}$, as follows
\begin{equation*}
    \mathbf{x} = \argmin_{\mathbf{x} \in \{\mathbf{x}^1, \ldots, \mathbf{x}^P\}} \sqrt{\sum_{i=1}^n \left(\bf{x}_{iA} - \bf{x}^{\rm o}_i\right)^2},
\end{equation*}
and this particular $\mathbf{x}$ is the trajectory that gets added into $\mathcal{D}$ altogether with its data-generating parameter. See also Algorithm \ref{algo:sl-abs-smc} which is discussed in section \ref{sec:abc-smc with bs}.}

\section{Dynamic ABC-SMC with data-conditional simulation}\label{sec:abc-smc with bs}
It remains that the ideas presented in the previous sections be combined to give a unified ABC-SMC framework for SDE inference. Dynamically, we learn the summary statistics with the PEN from Section \ref{sec:summaries}, which means summaries undergo re-learning in every ABC-SMC round, echoing the method in \cite{chen2020neural}. Prior to ABC-SMC, we produce a dataset comprising of prior-predictive samples, and train PEN on this dataset. At the initial round the summary statistics function is denoted as $S_1(\cdot)$. With every new round (say, round $t$), the dataset is expanded with newly accepted parameters and trajectories, and PEN is re-trained on this dataset, yielding an updated summary statistics function that we denote by $S_t(\cdot)$. 
As outlined in Algorithm \ref{algo:abs-smc}, a parameter $\bs{\theta}^*$ is proposed from the density $g_t(\cdot)$, and a summary statistic $\tilde{\bf{s}}$ is implicitly sampled from the data-conditional simulator $\tilde{\bf{s}}_t \sim p(\bf{s} \given \bf{x}^{\rm o}, \bs{\theta}^*)$ by sampling the conditional trajectory $\tilde{\bf{x}} \sim p(\bf{x} \given \bf{x}^{\rm o}, \bs{\theta}^*)$ and then computing $\tilde{\bf{s}}_t = S_t(\tilde{\bf{x}})$. The importance weight in this multi-round scenario takes the form
\begin{equation}
        w_t(\bs{\theta}, S_t(\tilde{\bf{x}})) \appropto \frac{\mathbbm{1}( \Vert S_t(\tilde{\bf{x}}) - \bf{s}_t^{\rm o}) \Vert < \epsilon_t)\pi(\bs{\theta})}{\sum_{i=1}^M W^i_{t - 1} \mathcal{N}(\bs{\theta} \given \bs{\theta}_{t- 1}^i, \bm{\Sigma}_{t - 1})} \frac{\mathcal{N}(S_t(\tilde{\bf{x}}) \given \bs{\mu}_{P, \bs{\theta}}, \bs{\Sigma}_{P, \bs{\theta}}) }{\mathcal{N}(S_t(\tilde{\bf{x}}) \given \tilde{\bs{\mu}}_{P, \bs{\theta}}, \tilde{\bs{\Sigma}}_{P, \bs{\theta}})} \label{eq:seq-imp-weight}.
\end{equation}
\begin{algorithm}[t]
    \caption{Dynamic ABC-SMC with data-conditional simulation {(ABC-SMC-DC)}}\label{algo:sl-abs-smc}

    \begin{algorithmic}[1]
    \STATE Initialize the dataset $\mathcal{D}=\emptyset$.
    \STATE Generate $R$ prior-predictive samples $(\mathbf{x}^{1:R}, \boldsymbol{\theta}^{1:R}) \overset{\text{iid}}{\sim} p(\mathbf{x} \given \boldsymbol{\theta}) \pi(\boldsymbol{\theta})$ and append $\mathcal{D} \coloneqq \mathcal{D} \cup (\mathbf{x}^{1:R}, \boldsymbol{\theta}^{1:R} )$.
    \STATE Train PEN on $\mathcal{D}$ to obtain $S_1(\cdot)$ and summarize the observation $\mathbf{s}^{\rm o}_1 = S_1(\mathbf{x}^{\rm o})$. 
    \FOR{$i = 1$ to $M$}
            \STATE Sample $\boldsymbol{\theta}^i_{{1}} \sim \pi(\boldsymbol{\theta})$ and sample a particle system $(\bf{x}^{1:P}_{1:N}, \omega^{1:P}_{1:N})$ with Algorithm \ref{algo:SIS}.
            \STATE Sample backward trajectory $\tilde{\bf{x}}$ with Algorithm \ref{algo:BSPS} and compute $d_1^i = \Vert S_1(\tilde{\bf{x}}) - \mathbf{s}^{\rm o}_1 \Vert$. 
            \STATE Store $\boldsymbol{\theta}^i_1$, compute the parameter weight $w_1^i$ by $\mathcal{N}(S_1(\tilde{\bf{x}}) \given \bs{\mu}_{P, \bs{\theta}}, \bs{\Sigma}_{P, \bs{\theta}}) / \mathcal{N}(S_1(\tilde{\bf{x}}) \given \tilde{\bs{\mu}}_{P, \bs{\theta}}, \tilde{\bs{\Sigma}}_{P, \bs{\theta}})$ using $(\bf{x}^{1:P}_{1:N}, \omega^{1:P}_{1:N})$.
            \STATE Pick a trajectory $\bf{x}$ from the forward trajectories $\bf{x}^{1:P}_{1:N}$ as detailed in Section \ref{sec:summaries}, and append it to the dataset $\mathcal{D} \coloneqq \mathcal{D} \cup (\mathbf{x}, \boldsymbol{\theta}^i_{{1}})$.
        \ENDFOR
        \FOR{$t = 2$ to $T$}
                \STATE Normalize $w_{t - 1}^{1:M}$.
                Retrain PEN  on $\mathcal{D}$ to obtain $S_{t}(\cdot)$ and summarize $\mathbf{s}^{\rm o}_t = S_{t}(\bf{x}^{\rm o})$.
            \STATE Compute particle covariance $\mathbf{\Sigma}_t = 2 \times {\rm Cov}((\bs{\theta}^{1:M}_{t - 1}, w_{t - 1}^{1:M}))$.
            \STATE Take $\epsilon_t$ to be the $\alpha$-quantile of distances $d_{t - 1}^{1:M}$ corresponding to accepted particles.
            \FOR{$i = 1$ to $M$}
                \WHILE{parameter not accepted}
                    \STATE Sample $\bs{\theta}^*$ from $\bs{\theta}_{t - 1}^{1:M}$ with probabilities $w_{t - 1}^{1:M}$ and perturb $\bs{\theta}^{i}_t \sim \mathcal{N}(\bs{\theta}^*, \mathbf{\Sigma}_t)$. 
                    \STATE Sample a particle system $(\bf{x}^{1:P}_{1:N}, \omega^{1:P}_{1:N})$ with Algorithm \ref{algo:SIS}.
                    \STATE Sample backward trajectory $\tilde{\bf{x}}$ with Algorithm \ref{algo:BSPS} and compute $d_t^i = \Vert S_t(\tilde{\mathbf{x}}) - \mathbf{s}^{\rm o}_t \Vert$.
                 \IF{$d_t^i \leq \epsilon_t$} 
                 \STATE Accept $\boldsymbol{\theta}^i_t$, compute the parameter weight $w_t^i$ by \eqref{eq:seq-imp-weight} using $(\bf{x}^{1:P}_{1:N}, \omega^{1:P}_{1:N})$.
                 \STATE Pick a trajectory $\bf{x}$ from the forward trajectories $\bf{x}^{1:P}_{1:N}$ as detailed in Section \ref{sec:summaries}, and append it to the dataset $\mathcal{D} \coloneqq \mathcal{D} \cup (\mathbf{x}, \boldsymbol{\theta}^i_{{t}} )$.
                        \ENDIF
                \ENDWHILE
            \ENDFOR
        \ENDFOR
        \STATE \textbf{Output:} Weighted sample $(\boldsymbol{\theta}^{1:M}_T, w_{T}^{1:M})$ of the ABC posterior distribution.
    \end{algorithmic}
\end{algorithm}
The complete procedure is given in Algorithm \ref{algo:sl-abs-smc}, {and our data-conditional inference method is denoted with ABC-SMC-DC, while we call ABC-SMC-F (where F stands for ``forward simulation'') a typical ABC-SMC procedure where simulated data is produced by the forward model}. We do not predefine the thresholding schedule $\epsilon_1 > \epsilon_2 > ... > \epsilon_T$ of ABC-SMC, but rather utilize the adaptive schedule proposed in \cite{prangle2017adapting}, where the threshold $\epsilon_{t + 1}$ is chosen as an $\alpha$-quantile of the distances corresponding to accepted summaries at round $t$.
Recall from end of section \ref{sec:summaries} that the training data $\mathcal{D}$ for PEN is combined in a different way from that of the standard ABC-SMC algorithm. Because PEN is designed to learn from samples from the forward model, and not from the data-conditional model, $\mathcal{D}$ cannot include the accepted trajectories $\tilde{\bf{x}}$ where $\tilde{\bf{x}} \sim p(\bf{x} \given \bf{x}^{\rm o}, \bs{\theta}^*)$. 

\section{Simulation studies}\label{sec:examples}
\subsection{Chan-Karaolyi-Longstaff-Sanders family of models}
The Chan-Karaolyi-Longstaff-Sanders (CKLS) family is a class of parametric SDE models that are widely used in finance applications, in particular to model interest rates and asset prices \citep{chan1992empirical}. The diffusion term of CKLS depends on two parameters and the state, taking the form $\sigma X_t^{\gamma}$, where $\sigma>0$ is the diffusion coefficient and $\gamma\in[0,1]$. This state-dependent diffusion term can substantially increase the randomness of the SDE paths, and therefore CKLS presents a considerable inferential challenge for ABC algorithms when the forward simulator is used. The CKLS process satisfies the SDE 
\begin{equation}
    \begin{cases}
            {\rm d}X_t = \beta (\alpha - X_t) {\rm d}t + \sigma X_t^{\gamma} {\rm d}B_t,  & \text{if } t > 0, \\
        X_0 = x_0, & \text{if } t = 0,
    \end{cases}
    \label{eq:ckls}
\end{equation}
for $\alpha, \beta \in \mathbb{R}$ and $\sigma, \gamma \in \mathbb{R}_+$. The CKLS family includes the Ornstein--Uhlenbeck process for $\gamma = 0$, the Cox--Ingersoll--Ross process for $\gamma = 1/2$, and the Black--Scholes process for $\alpha = 0$ and $\gamma = 1$. We will restrict ourselves to the case where $\alpha, \beta > 0$ and $0 \leq \gamma < 1$. This section describes numerical experiments for the different SDE models from the CKLS family, with tractable as well as intractable likelihoods. For the CKLS model, we already illustrated the benefits of generating trajectories by the data-conditional simulator in Figure \ref{fig:ckls-trajectories-true-and-pps}. A similar figure for the Ornstein--Uhlenbeck, Cox-Ingersoll-Ross and an SDE model with nonlinear drift, can be seen in the Supplementary Material. We evaluate the efficiency of our proposed Algorithm \ref{algo:sl-abs-smc} compared to ABC-SMC with Euler-Maruyama as a forward simulator. For the data-conditional simulator we will take as weighing functions the Gaussian densities induced by the EM approximation of \eqref{eq:ckls}, namely, 
\begin{equation}
    q(x^{\rm o}_i \given x_{iA - 1}) = \mathcal{N}(x^{\rm o}_i \given x_{iA - 1} +  \beta(\alpha - x_{iA - 1}) h, \sigma^2 x_{iA - 1}^{2 \gamma}  h). \label{eq:ckls-gaussian}
\end{equation} 

We run Algorithms \ref{algo:abs-smc} and \ref{algo:sl-abs-smc} with $M = 10,000$ parameter particles and $T = 20$ rounds ($T = 10$ for the simpler case of the Ornstein--Uhlenbeck), but we preemptively stop the inference once the acceptance rate of the parameters falls below 1.5\ when $t > 2$. In our experiments, we use the Wasserstein distance as a measure of similarity between probability distributions. If the true posterior is available, we use that as a reference posterior. In the case of an intractable likelihood, we choose as reference the ABC posterior obtained from a run of the standard ABC-SMC algorithm with the same stopping condition and with fixed summary statistics, obtained by training PEN on 300,000 prior-predictive samples before starting ABC-SMC. {Throughout this subsection, PEN is composed of dense layers and is structured as 2-100-100-100-dim($\bs{\theta}$).}
\subsubsection{Fixed $\gamma = 0$ (Ornstein-Uhlenbeck): inference for \texorpdfstring{$(\alpha, \beta, \sigma)$}{TEXT}} 
\begin{figure}[tp]
    \centering
    \includegraphics[scale=1.2]{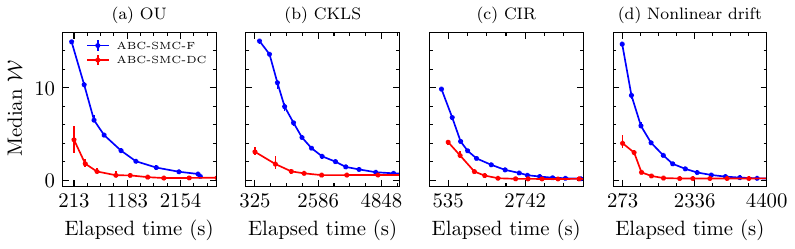}
    \caption{Results from (a) Ornstein--Uhlenbeck, (b) Chan--Karaolyi--Longstaff--Sanders, (c) Cox--Ingersoll--Ross and (d) a SDE with Nonlinear drift. The median Wasserstein distance between the ABC-posteriors at each round and a reference posterior is shown on the y-axis, and the median cumulative elapsed time per round is shown on the x-axis. A circle represents a round of ABC-SMC. Medians are taken over 20 runs. Blue represents the standard ABC-SMC algorithm with forward simulation, and red represents Algorithm \ref{algo:sl-abs-smc}.}
    \label{fig:wassersteins}
\end{figure}

In this example, the CKLS model is considered with $\gamma$ fixed to \(\gamma = 0\), yielding the ubiquitously utilized Ornstein--Uhlenbeck (OU) process. A notable characteristic of the OU process is that its transition densities are known. Specifically, the transition density from state \(x_s\) to state \(x_t\), for \(s < t\), is expressed as
\begin{equation}
    p(x_t \given x_s) = \mathcal{N}(x_t \given \alpha + (x_s - \alpha) e^{-\beta (t - s)}, \sigma^2 (1 - e^{-2 \beta (t - s))}) / 2\beta). \label{eq:ou-tdensity}
\end{equation}
To this end, as data $\bf{x}^{\rm o}$, a discrete realization of length $100$ is generated with an initial state \(x_0 = 0.01\) and parameters \(\boldsymbol{\theta}_{\rm true} = (\alpha , \beta, \sigma) = (3, 1, 1)\) by sampling from the exact transition densities \eqref{eq:ou-tdensity} with \(\Delta t = t-s =0.1\). For both the forward and data-conditional simulator, the number of subintervals between the observations is set to $A = 10$, hence the integration timestep is $h = \Delta t / A = 0.01$. The number of trajectories for the data-conditional simulator is set to $P = 30$. Despite the availability of the transition densities, to mimic an intractable scenario we use \eqref{eq:ckls-gaussian} as weighing function  with $\gamma = 0$. 
We now proceed to parameter inference for the OU process, treating \(\gamma\) as known and fixed at \(\gamma = 0\), while \((\alpha, \beta, \sigma)\) are unknown. The OU process is an example of a tractable diffusion because the likelihood function is available via a product of transition densities whose form is given by \eqref{eq:ou-tdensity}. Therefore exact inference via MCMC is possible. We define an independent prior distribution over the parameters \((\alpha, \beta, \sigma)\) by specifying \(\pi(\alpha, \beta, \sigma) = \mathcal{U}(\alpha \given 0, 30) \mathcal{U}(\beta \given 0, 10) \mathcal{U}(\sigma \given 0, 2)\), where \(\mathcal{U}(a \given b, c)\) denotes a uniform distribution with bounds \(b\) and \(c\) for parameter \(a\). To obtain an initial estimate of the summary statistics, we pretrain PEN on $R = 20,000$ prior-predictive samples. {Our data-conditional ABC-SMC is denoted with ABC-SMC-DC, while pure forward simulation is denoted with ABC-SMC-F.}
The result of the analysis is in Figure \ref{fig:wassersteins} (a) and is based on 20 independent runs on the same observed trajectory $\bf{x}^{\rm o}$. The figure displays the computation time versus the Wasserstein distance between the ABC posteriors and the reference posterior (obtained via MCMC using the exact likelihood determined by \eqref{eq:ou-tdensity}). In this case, the parameters of the OU model are easier to estimate compared to the general CKLS instance examined further below, where all parameters are unknown and the diffusion is state-dependent. 
However even in this simpler case, Figure \ref{fig:wassersteins}(a) clearly displays that with our sampler we require about 3-4 rounds, and about 8-9 rounds with the standard sampler. {Reducing the total number of rounds is important in our context, as at each round the PEN network is retrained, which brings a consequent computational overhead.} In terms of running time, by comparing similar Wasserstein distances that are obtained at round 3 with our sampler and at round 8 with the standard sampler, it takes about 600 seconds to reach round 3 in the former case and around 2,100 seconds to reach round 8 for the latter, thus with our sampler we have a 3.5-fold acceleration on average. In Table \ref{tab:nonlin-table} we show that by round 3-4 the data-conditional ABC-SMC achieves a significant reduction of the Wasserstein distance, while showing a larger acceptance rate than the pure forward approach. For later rounds, the acceptance rate of our simulator decreases, however this is expected since very small Wasserstein distances are achieved.  

\begin{table}[h]
\centering
\scalebox{0.8}{\begin{tabular}{|c|c|c|c|c|c|c|c|c|c|c|}
\hline
\textbf{Round} & \textbf{1} & \textbf{2} & \textbf{3} & \textbf{4} & \textbf{5} & \textbf{6} & \textbf{7} & \textbf{8} & \textbf{9} & \textbf{10} \\
\hline 
\multicolumn{11}{|c|}{\textbf{CKLS SDE with $(\alpha, \beta, \sigma)$ unknown and $\gamma = 0$ (Ornstein--Uhlenbeck)}} \\
\hline
Acc. rate (F/B) & $100 / 100$ & $46 / 42$ & $37 / 42$ & $29 / 30$ & $21 / 30$ & $18 / 20$ & $17 / 11$ & $12 / 7$ & $10 / 5$ & $8 / 2$ \\
\hline
Wasser. (F/B) & $15 / 4.4$ & $10 / 1.8$ & $6.5 / 0.9$ & $4.9 / 0.5$ & $3.2 / 0.5$ & $2 / 0.3$ & $1.3 / 0.2$ & $0.9 / 0.2$ & $0.7 / 0.3$ & $0.4 / 0.3$ \\
\hline
\multicolumn{11}{|c|}{\textbf{CKLS SDE with $(\alpha, \beta, \sigma, \gamma)$ unknown}} \\
\hline 
Acc. rate (F/B) & $100 / 100$ & $54 / 52$ & $39 / 19$ & $26 / 19$ & $22 / 5$ & $18 / 2$ & $14 / 1$ & $12 / 1$ & $9 / 1$ & $6 / \text{NA}$ \\
\hline
Wasser. (F/B) & $15 / 3$ & $13.6 / 1.7$ & $10.5 / 0.9$ & $8 / 0.7$ & $6.2 / 0.5$ & $4.6 / 0.5$ & $3.5 / 0.5$ & $2.6 / 0.5$ & $2 / 0.6$ & $1.4 / \text{NA}$ \\
\hline
\multicolumn{11}{|c|}{\textbf{CKLS SDE with $(\alpha, \beta, \sigma)$ unknown and $\gamma = 1 / 2$ (Cox--Ingersoll--Ross)}} \\
\hline
Acc. rate (F/B) & $100 / 100$ & $54 / 78$ & $34 / 50$ & $31 / 55$ & $22 / 40$ & $17 / 33$ & $15 / 25$ & $13 / 14$ & $11 / 8$ & $9 / 5$ \\
\hline
Wasser. (F/B) & $9.8 / 4.1$ & $6.8 / 2.7$ & $4.2 / 0.9$ & $3.2 / 0.5$ & $2.3 / 0.2$ & $1.6 / 0.1$ & $1.1 / 0.1$ & $0.8 / 0.1$ & $0.5 / 0.1$ & $0.4 / 0.1$ \\
\hline
\multicolumn{11}{|c|}{\textbf{SDE with nonlinear drift}} \\
\hline
Acc. rate (F/B) & $100 / 100$ & $50 / 85$ & $36 / 53$ & $30 / 59$ & $25 / 46$ & $21 / 37$ & $18 / 24$ & $15 / 14$ & $13 / 8$ & $11 / 4$ \\
\hline
Wasser. (F/B) & $14.7 / 4$ & $9.2 / 3$ & $5.9 / 0.8$ & $4 / 0.5$ & $2.7 / 0.2$ & $1.8 / 0.2$ & $1.2 / 0.2$ & $0.8 / 0.2$ & $0.5 / 0.2$ & $0.4 / 0.2$ \\
\hline
\end{tabular}}
\caption{Comparison of median acceptance rates (in \%) and median Wasserstein distances for both ABC-SMC algorithms, forward (F) and forward-backward data-conditional simulation (B), in the first 10 rounds. Each cell in the table reports numbers as $a/b$, where $a$ refers to the performance using F and $b$ refers to B.}
\label{tab:nonlin-table}
\end{table}

\subsubsection{CKLS with four unknown parameters \texorpdfstring{$(\alpha, \beta, \sigma, \gamma)$}{TEXT}} \label{sec:ckls-intractable}
\begin{figure}[tp]
    \centering
    \includegraphics[scale=0.9]{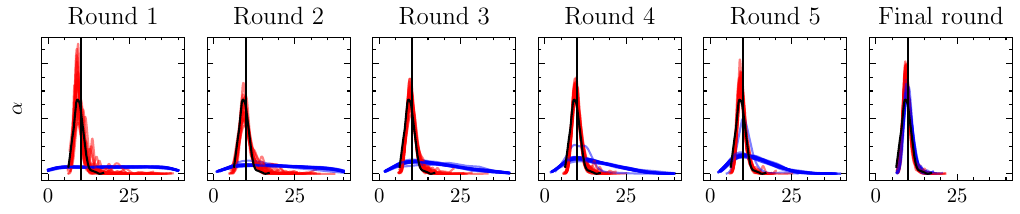}
    \includegraphics[scale=0.9]{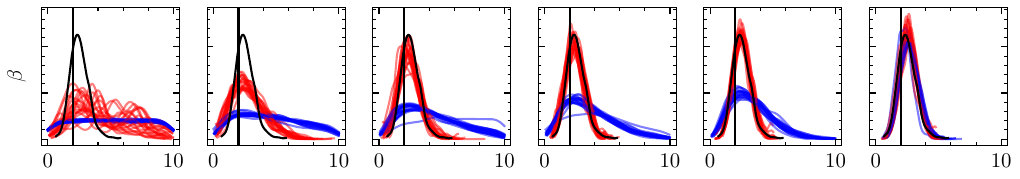}
    \includegraphics[scale=0.9]{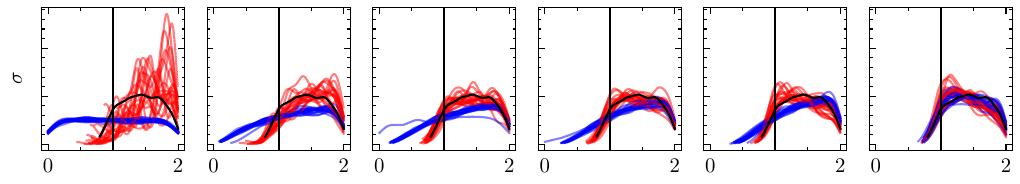}
    \includegraphics[scale=0.9]{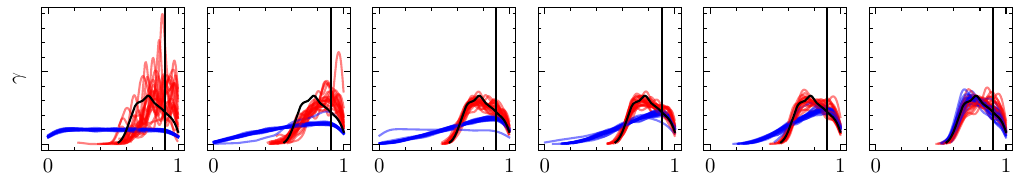}
    \caption{CKLS with four unknown parameters. Marginal posterior distributions across several rounds (rounds) of ABC-SMC: marginals from our data-conditional ABC-SMC-B are in red, and those from the forward ABC-SMC-F are in blue. The reference marginal posteriors are depicted in black, and the true parameter values as black vertical lines.}
    \label{fig:ckls-posteriors-row}
\end{figure}
In this example, we examine the SDE \eqref{eq:ckls}, where the parameters \((\alpha, \beta, \sigma, \gamma)\) are unknown, and the likelihood is intractable. A discrete realization of the CKLS model is generated with an initial state of \(x_0 = 0.1\) and true parameters \(\boldsymbol{\theta}^{\rm true} = (\alpha, \beta, \sigma, \gamma) = (10, 2, 1, 0.9)\). This is achieved by numerically solving the SDE \eqref{eq:ckls} using the EM method and an integration timestep of \(\Delta t = 0.0001\). After generating 100,000 data points, every 1,000th point is retained, resulting in a sample of length \(n = 100\) at timepoints \(0, 0.1, \ldots, 10\). This set of \(n\) observations is denoted as our data \(\mathbf{x}^o\). For both the forward and data-conditional simulator, the number of subintervals between the observations is set to $A = 10$, hence the integration timestep is $h = \Delta t / A = 0.01$. The number of trajectories for the data-conditional simulator is set to $P = 50$. Figure \ref{fig:ckls-trajectories-true-and-pps} displays the observed data alongside simulated trajectories from both simulators.
We set independent uniform priors over the parameters \((\alpha, \beta, \sigma, \gamma)\) by specifying \(\pi(\alpha, \beta, \sigma, \gamma) = \mathcal{U}(\alpha \given 0, 40) \mathcal{U}(\beta \given 0, 10) \mathcal{U}(\sigma \given 0, 2) \mathcal{U}(\gamma | 0, 1)\). To obtain an initial estimate of the summary statistics, we pretrain PEN on $R = 50,000$ prior-predictive samples. The analysis results are illustrated in Figure \ref{fig:wassersteins}(b), derived from 20 independent runs on the observed trajectory depicted in Figure \ref{fig:ckls-trajectories-true-and-pps}. Additionally, the posterior distributions resulting from these 20 runs are visualized in Figure \ref{fig:ckls-posteriors-row}. 
Figure \ref{fig:wassersteins} reveals a significant reduction in the Wasserstein distance even from the initial round using our method, as opposed to the standard ABC-SMC-F. Moreover, when viewed in conjunction with Figure \ref{fig:ckls-posteriors-row}, it is evident that our method attains satisfactory inference for all parameters by the third round. ABC-SMC-F takes roughly 2,500 seconds to reach a Wasserstein distance of 3.1, while the data-conditional ABC-SMC-DC starts at that value already at round one. Additionally, ABC-SMC needs about 4,645 seconds to attain a Wasserstein distance of 0.9, compared to our method, which requires 1,625 seconds to reach the same distance, thus with our sampler we have a 2.8-fold acceleration on average. See Table \ref{tab:nonlin-table} for further results: there it is clear that at round 2 the large decrease in the Wasserstein distance brought by our approach is not paired with a drastic drop in the acceptance rate, since the latter is on par with the pure forward simulator. This is a strength of our simulator and, as previously mentioned, the run of our data-conditional simulator here could be halted at round 3.
\subsection{Biochemical reaction networks}

{
A biochemical reaction network consists of a set of $d$ chemical species, $X_1, \ldots, X_d$ that interact via a network of $R$ reactions
\begin{equation}
    \mathcal{R}_j: \sum_{i=1}^d \nu_{i,j}^- X_i \xrightarrow{\theta_j} \sum_{i=1}^d \nu_{i,j}^+ X_i, \quad j = 1, 2, \ldots, R,
\end{equation}
where $\nu_{i,j}^-$ and $\nu_{i,j}^+$ are the number of reactant and product molecules. When the system contains a large number of molecules and reactions occur frequently, the behavior of the biochemical reaction network can be closely approximated using the chemical Langevin equation \citep[see e.g.][]{wilkinson2018stochastic}. The chemical Langevin equation is an It\^{o} SDE of the form
\begin{equation}
    {\rm d}\mathbf{X}_t = \sum_{j=1}^R \boldsymbol{\nu}_j a_j(\mathbf{X}_t) {\rm d}t + \sum_{j=1}^R \boldsymbol{\nu}_j \sqrt{a_j(\mathbf{X}_t)} {\rm d}\mathbf{B}_t^{(j)} 
    \label{eq:cle}
\end{equation}
where $\boldsymbol{\nu}_j$ is the $j$th column of the stoichiometry matrix $\boldsymbol{\nu}$ with elements $\nu_{i,j} = \nu_{i,j}^+ - \nu_{i,j}^-$, $a_j(\cdot)$ is the hazard of reaction $j$ (typically computed via a mass-action rate law) and $\mathbf{B}_t^{(j)}$, $j=1\ldots,R$, are uncorrelated Brownian motion processes. For these models the parameter inference problem involves calculating the posterior distribution of the reaction rate constants $\theta_1, \ldots, \theta_R$. }
{
This section presents numerical experiments for two distinct types of biochemical reaction networks: the Schl{\"o}gl model and the Lotka--Volterra model. The Schl{\"o}gl model is a one-dimensional SDE that demonstrates stochastic bistability. The Lotka-Volterra model is a two-dimensional SDE and is explored for its characteristic oscillatory behavior. For both models we generate an observation by numerically solving the respective SDE via the EM method with integration timestep of $\Delta t = 0.0001$, on the interval $[0, 50]$. After generating 500,000 data points we retained every 10,000th point and thus obtained an observation at the timepoints $0, 1, \ldots, 50$. We configure both the forward and data-conditional simulator to have $A = 100$ subintervals between observations. Additionally, we enhanced the complexity of PEN by adding three 1D convolutional layers (25-50-100) before the three dense layers (100-100-100). This is due to the previous architecture's inability to accurately learn the summaries for these more complex models. Additionally, using the Gaussian density on the smaller interval $[t - h, t]$ as a weighting function, similar to what was done in \eqref{eq:ckls-gaussian}, is not feasible because it turned out to be excessively narrow and stringent in particle selection when applied to the chemical Langevin equation \eqref{eq:cle}. To address this issue, we construct a specific weighting function for both models considered in this subsection. The algorithms are ran with $M = 10000$ particles, for a maximum of $T = 20$ rounds or until an acceptance rate smaller than $1.5\%$ is achieved.}

\subsubsection{Lotka--Volterra model}

\begin{figure}[tp]
    \centering
    \includegraphics[scale=0.95]{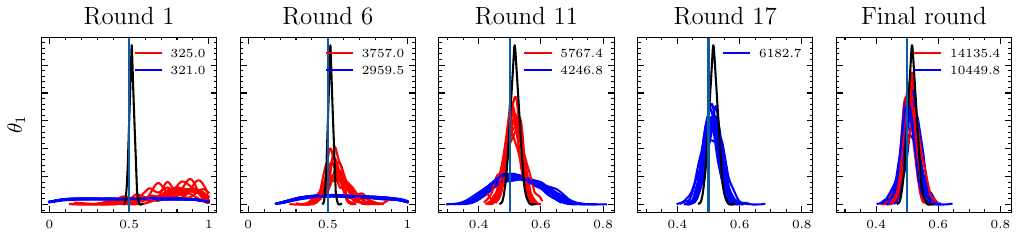}
    \includegraphics[scale=0.95]{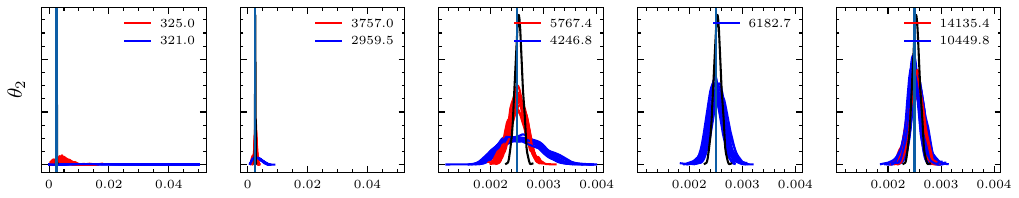}
    \includegraphics[scale=0.95]{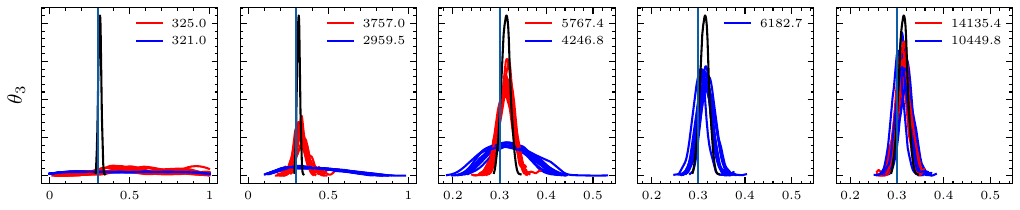}
    \caption{{{Lotka--Volterra model. Marginal posterior distributions at several rounds of ABC-SMC: for ABC-SMC-DC the marginals are displayed in red, and for ABC-SMC-F in blue (ABC-SMC-DC terminated before round 17). The reference marginal posteriors are depicted in black, and the true parameter values as black vertical lines. Each panel reports, in the upper right corner, the number of seconds since the algorithm started.}}}
    \label{fig:lv-posteriors}
\end{figure}

{
The Lotka--Volterra model is composed of two biochemical species, predator and prey, and three reactions (prey reproduction, prey death and predator reproduction, predator death). The reactions and their propensities are given as follows
\begin{equation}
    \mathcal{R}_1: X \xrightarrow{\theta_1} 2X, \quad \mathcal{R}_2: X + Y \xrightarrow{\theta_2} 2Y, \quad \mathcal{R}_3: Y \xrightarrow{\theta_3} \emptyset.
\end{equation}
Let \(\mathbf{X}_t = (X_t, Y_t)\) denote the populations of prey \(X_t\) and predators \(Y_t\) at any given time \(t\). The stoichiometry associated with the system is 
\begin{equation}
    \boldsymbol{\nu} = 
    \begin{pmatrix}
        1 & -1 & 0 \\
        0 & 1 & -1
    \end{pmatrix}.
\end{equation}
The model is governed by the following SDE (see \citealp{golightly2011bayesian} for details)
\begin{equation}
        {\rm d} \begin{pmatrix}
        X_t \\ 
        Y_t
    \end{pmatrix} = 
    \begin{pmatrix}
        \theta_1 X_t - \theta_2 X_t Y_t \\
        \theta_2 X_t Y_t - \theta_3 Y_t 
    \end{pmatrix}
    {\rm d}t + 
    \begin{pmatrix}
        \sqrt{\theta_1 X_t} & -\sqrt{\theta_2 X_t Y_t} & 0 \\
        0 & \sqrt{\theta_2 X_t Y_t} & -\sqrt{\theta_3 Y_t}
    \end{pmatrix}
    {\rm d}
    \begin{pmatrix}
        B^{(1)}_t \\
        B^{(2)}_t \\
        B^{(3)}_t
    \end{pmatrix}.
    \label{eq:lv-sde}
\end{equation}
We generate a discrete realization of \eqref{eq:lv-sde} with initial state $\mathbf{x}^{\rm o}_0 = (100, 100)^T$ and parameters $\boldsymbol{\theta}^{\rm true} = (\theta_1, \theta_2, \theta_3) = (0.5, 0.0025, 0.3)$. For the data-conditional simulator we set the number of trajectories to $P = 30$. We adopt an independent prior specification by taking $\pi(\boldsymbol{\theta}) = \mathcal{U}(\theta_1 \given 0, 1) \mathcal{U}(\theta_2 \given 0, 0.05) \mathcal{U}(\theta_3 \given 0, 1)$. As a weighting function we take the multivariate Gaussian density induced by the EM approximation, however, we rescale the covariance by $\Delta / (2h)$. Namely, The weighting function is a multivariate Gaussian density with mean
\begin{equation}
    \begin{pmatrix}
        x_{iA - 1} \\ y_{iA - 1}
    \end{pmatrix}
    + h 
    \begin{pmatrix}
    \theta_1 x_{iA - 1} - \theta_2 x_{iA - 1} y_{iA - 1} \\
        \theta_2 x_{iA - 1} y_{iA - 1} - \theta_3 y_{iA - 1}
    \end{pmatrix}
    \label{eq:em-mean}
\end{equation}
and covariance 
\begin{equation}
    \frac{\Delta t}{2}\begin{pmatrix}
        \theta_1 x_{iA - 1} + \theta_2 x_{iA - 1} y_{iA - 1} & - \theta_2 x_{iA - 1} y_{iA - 1} \\
        -\theta_2 x_{iA - 1} y_{iA - 1} & \theta_3 y_{iA - 1} + \theta_2 x_{iA - 1} y_{iA - 1}
    \end{pmatrix}.
    \label{eq:lv-covariance}
\end{equation}
This adjustment prevents the weighting function from being excessively narrow and stringent in particle selection. We included a stopping rule to fix the summary statistics function and halt neural network training. At each round, a new dataset is obtained, expanding the training and validation sets. If the current model (round $t$) performs worse or only marginally better on the expanded validation set than the previous model (round $t-1$) tested on this same validation set, training is stopped, and the summary statistics remain unchanged. 
A reference posterior was returned by a run of (correlated particles) pseudomarginal Metropolis-Hastings, with the likelihood function obtained using the modified diffusion bridge of Durham and Gallant (the version found in \citealp{whitaker2017}) to impute points between observations. 
The results of the analysis can be seen in Figure \ref{fig:lv-posteriors}. We notice that for $\theta_2$, ABC-SMC-DC finds a region of high posterior density in the first round (318 seconds on average), whereas for ABC-SMC-F about 6 rounds are required (3206 seconds on average) to arrive at a similar posterior approximation. For ABC-SMC-DC, a crude posterior approximation is already obtained by round 6 with an average Wasserstein distance of $0.08$ ($0.28$ for ABC-SMC-F), and displaying an inference quality that by round 11 is considerably more accurate than ABC-SMC-F, with an average Wasserstein distance of $0.015$ ($0.07$ for ABC-SMC-F). We recall that a possible use of ABC-SMC-DC is to rapidly find the bulk of the posterior (say in six rounds in this case), and from there initialize the standard ABC-SMC-F. 
}

\subsubsection{Schl{\"o}gl model}

\begin{figure}
    \centering
    \includegraphics[scale=0.95]{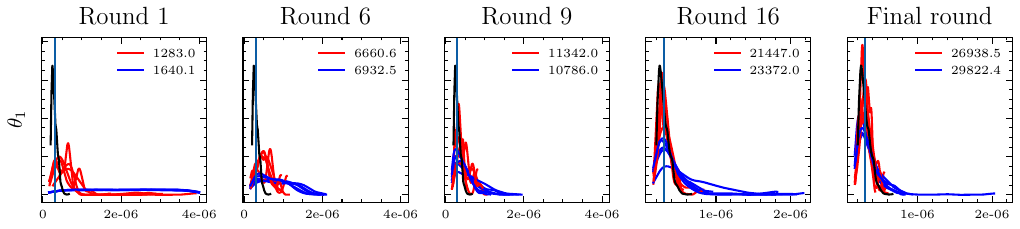}
    \includegraphics[scale=0.95]{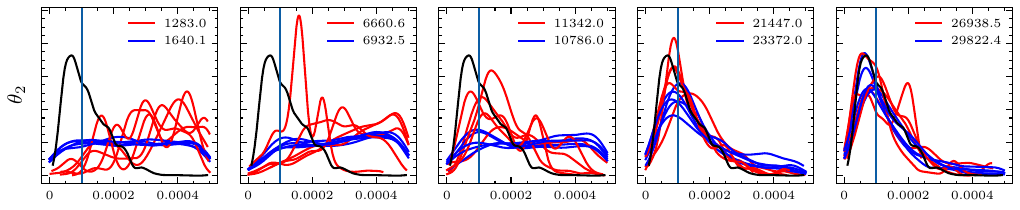}
    \includegraphics[scale=0.95]{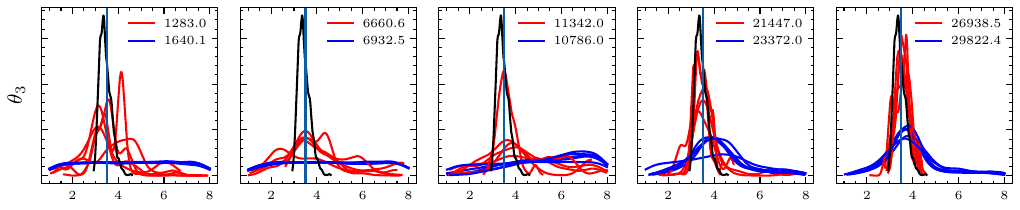}
    \caption{{Schl{\"o}gl model. Marginal posterior distributions for ABC-SMC-DC displayed in red, and ABC-SMC-F in blue. The reference marginal posteriors (via particle MCMC) are depicted in black, and the true parameter values as black vertical lines. Each panel reports, in the upper right corner, the number of seconds since the algorithm started.}}
    \label{fig:schlogl-posterior}
\end{figure}

\begin{figure}
    \centering
    \includegraphics[scale=1.3]{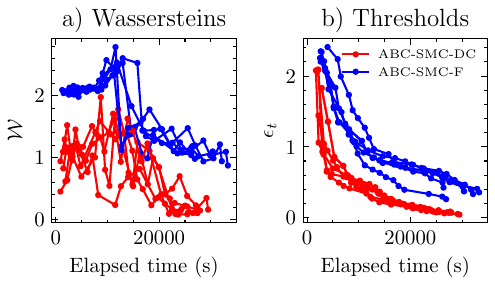}
    \caption{{Results from the Schl{\"o}gl model. 
    a) the Wasserstein distance between the ABC-posteriors at each round and the reference posterior from particle MCMC is shown on the y-axis, and the cumulative elapsed time per round is shown on the x-axis. b) the ABC thresholds are shown on the y-axis, and the x-axis is the same as for a). A circle represents a round of ABC-SMC. Blue represents the standard ABC-SMC algorithm with forward simulation, and red represents Algorithm \ref{algo:sl-abs-smc}.}}
    \label{fig:schlogl-wassersteins}
\end{figure}

The Schl{\"o}gl model is a notable instance of a reaction network showcasing bistability. It describes how, by a specific set of reactions, solutions of the deterministic representation gravitate towards one of two stable states and remain there indefinitely. In contrast, in stochastic versions of the model, the system can spontaneously alternate between these stable states due to inherent noise. The interval between switching events is unpredictable. This makes data from a single experiment challenging to replicate with simulated data in a particular experimental context, typically resulting in very low acceptance rates for ABC schemes. The reactions governing the Schlögl system are given as follows: 
\begin{equation}
    \mathcal{R}_1: A + 2X \xrightarrow{\theta_1} 3X, \quad \mathcal{R}_2: 3X \xrightarrow{\theta_2} A + 2X, \quad \mathcal{R}_3: B \xrightarrow{\theta_3} X, \quad  \mathcal{R}_4: X \xrightarrow{\theta_4} B.
\end{equation}
The stoichiometry associated with the system is $\boldsymbol{\nu} = (1, -1, 1, -1)$. In the following, we assume that the molecular numbers of the species $A$ and $B$ are fixed at constant values, $A = 10^5$, and $B = 2 \times 10^5$. The model is governed by the following SDE 
\begin{equation}
\begin{cases}
    \mathrm{d}X_t = (\theta_1 A X_t (X_t - 1)/2 - \theta_2 X_t(X_t - 1)(X_t - 2) / 6 + \theta_3 B - \theta_4 X_t) {\rm d}t \\
    \quad \quad + \sqrt{\theta_1 A X_t (X_t - 1)/2} \, \mathrm{d}B^{(1)}_t - \sqrt{\theta_2 X_t(X_t - 1)(X_t - 2) / 6}\, \mathrm{d}B^{(2)}_t  \\ \quad \quad + \sqrt{\theta_3 B}\, \mathrm{d}B^{(3)}_t - \sqrt{\theta_4 X_t} \, \mathrm{d}B^{(4)}_t, \\
    X_0 = x_0. 
\end{cases}
\label{eq:schlogl-sde}
\end{equation}
We generate a discrete realization of \eqref{eq:schlogl-sde} with initial state $x_0 = 249$ and parameters $\boldsymbol{\theta}^{\rm true} = (\theta_1, \theta_2, \theta_3, \theta_4) = (\num{3e-7}, \num{1e-4}, \num{1e-3}, 3.5)$. For the data-conditional simulator we set the number of trajectories to $P = 30$. We adopt an independent prior specification by taking $\pi(\boldsymbol{\theta}) = \mathcal{U}(\theta_1 \given \num{1.6e-7}, \num{4e-6}) \mathcal{U}(\theta_2 \given 0.0, \num{5e-4}) \mathcal{U}(\theta_4 \given 1, 8)$, but keeping $\theta_3$ fixed to $\num{1e-3}$. As a weighting function, we use the Gaussian density induced by the EM approximation; however, unlike \eqref{eq:ckls-gaussian} which is on $[t - h, t]$, we apply it over the interval $[t - 5h, t]$. To obtain an initial estimate of the summary statistics, we pretrain PEN on $R = 100,000$ prior-predictive samples. We obtained a reference posterior via particle MCMC on the original observation corrupted with a small amount of Gaussian measurement noise (with zero mean and standard deviation 15), using 5,000 particles and advanced using Metropolis random walk proposals, producing a good mixing and an acceptance rate of 22\%. Both ABC-SMC-F (Algorithm \ref{algo:abs-smc}) and ABC-SMC-DC (Algorithm \ref{algo:sl-abs-smc}) were run with a threshold selection $\alpha$-quantile of 0.75, because the previous value of 0.5 was too aggressive, often causing the algorithm to get stuck in a local optimum. Additionally, the stopping rule to fix the summary statistics function could not be applied in this example because the observed summary statistics failed to stabilize, leading to inaccurate inferences, even though the mean squared error (MSE) on the validation set for PEN did stabilize. This issue is likely due to the high variability in the Schl{\"o}gl model's paths, caused by the cubic and quadratic terms in the diffusion function, as well as its bistable behavior. To this end, if one is interested in fixing the summary statistics after some rounds, a balance should be achieved between the stability of the MSE on the continually evolving validation set and the stability of the observed summary statistics. Ultimately, the accept/reject step relies on the observed summary statistics being accurate and stable. The Wasserstein results are presented in Figure \ref{fig:schlogl-wassersteins}(a), and the ABC thresholds are shown in Figure \ref{fig:schlogl-wassersteins}(b), based on five independent runs. Additionally, the posterior distributions from these five runs are visualized in Figure \ref{fig:schlogl-posterior}. ABC-SMC-DC effectively localizes the posterior region for $\theta_1$ even in the first round. Conversely, ABC-SMC-F fails to find an accurate approximation of the $\theta_3$ posterior within the given time. ABC-SMC-DC achieves a much lower Wasserstein distance, and the ABC thresholds decrease much faster compared to ABC-SMC-F. This observation suggests a hybrid algorithm: once a sufficiently low threshold/Wasserstein distance is achieved, reverting to ABC-SMC-F can increase the inference accuracy at a lower computational cost.  

\section{Discussion}

We have constructed a novel, efficient approach for parameter inference in stochastic differential equation (SDE) models when using approximate Bayesian computation (ABC) algorithms. Our ABC method with \textit{data-conditional simulation} proposes trajectories resulting from smoothing approaches (using a carefully constructed combination of forward- and backward-simulated paths), producing a  method converging much more rapidly to some reference posterior, as shown via Wasserstein distances. This is especially evident in the first round of our ABC-SMC method, where the initial Wasserstein distance is much lower compared to the one produced from standard ABC-SMC where trajectories are instead myopically resulting from a forward model simulation. This often means that already at rounds 3--5 of our ABC-SMC the posterior approximation is close to the bulk of the reference posterior, and moreover having to run very few rounds is a favorable feature, given that at each round we also retrain a neural network to learn summary statistics, which is further discussed below. This could be exploited to create a ``hybrid'' algorithm, where our method is used for a few initial rounds allowing rapid converging to the bulk of the posterior, to then revert to purely forward ABC-SMC simulation for the remaining rounds, as the latter is computationally favorable when the approximated posterior is close to the target.

Our method was tested on simulation studies that include several members of the Chan-Karaolyi-Longstaff-Sanders family of SDEs, and we found that the most dramatic display of the efficiency of our method comes from the more complex case studies, where the solution to the SDE is very erratic. We conjecture that more drastic accelerations can be achieved with even more challenging case studies with very erratic solution paths.

An important component of our proposed ABC-SMC-DC scheme with backward simulation is that the forward-pass in the procedure can be executed with any numerical discretization scheme, not just the customary Euler-Maruyama scheme. While in the present work we have illustrated our approach by using Euler-Maruyama in step 4 of the lookahead sampler (Algorithm \ref{algo:SIS}), this need not be the case. Any higher-order scheme can be used in the forward-pass, and while this may imply that the scheme does not enjoy a closed-form expression for the associated transition density, the weights $\omega$ computed in step 4 of Algorithm \ref{algo:BSPS} are only necessary towards obtaining a backward trajectory. Moreover, these weights could be computed with an approximate transition density such as the one resulting from the Euler-Maruyama or the Milstein schemes. This apparent contradiction means that, although when using weights $\omega$ from the latter two schemes to pick in the backward pass a trajectory from the cloud generated in the forward pass (which would instead use a higher-order method) we would have an inconsistency in the construction of the $\omega$-weights in Algorithm \ref{algo:BSPS}, nevertheless this will allow to sample a backward trajectory that is picked from a set of accurate forward trajectories. This procedure could still be highly beneficial, as \cite{buckwar2020spectral} have shown that the ABC posterior can be inaccurate when employing the Euler-Maruyama scheme, for some SDE models.

Another important feature of our ABC-SMC with data-conditional simulation is that it makes use of sequential learning of the summary statistics, which are refined after each ABC-SMC round, using the neural-network denoted PEN \citep{wiqvist2019partially}, which is especially suited to Markov processes. However, we wish to emphasize that while PEN is a recommended choice to automatically construct summary statistics for SDEs, since PEN is by construction designed to exploit the Markovianity in the paths of SDEs solutions, the user may decide to use other ways to determine the summary statistics while still using our Algorithm \ref{algo:sl-abs-smc}. While this is entirely possible, the user should be careful in doing so, as in such case the synthetic likelihoods appearing in \eqref{eq:seq-imp-weight} may not be an appropriate approximation to the density of the summary statistics. The latter aspect is less risky when the summary statistic is an approximation to the parameters posterior mean (as with PEN, but also with the linear regression method of \citealp{fearnhead2012constructing}, and the neural-network approaches of \citealp{jiang2017learning} and \citealp{akesson2021convolutional}), thanks to central limit theorem arguments. In fact we only needed 30-50 paths to approximate the synthetic likelihoods via PEN. While PEN can so far be used only with one-dimensional SDEs, this does not prevent the use of our backwards simulation ABC-SMC with multidimensional SDEs, by constructing summary statistics using alternative methods, such as the already mentioned methods of \cite{fearnhead2012constructing}, \cite{jiang2017learning} or \cite{akesson2021convolutional}.

\section*{Acknowledgments}
UP acknowledges funding from the Swedish National Research Council (Vetenskapsrådet 2019-03924). PJ and UP acknowledge funding from the Chalmers AI Research Centre. We would like to express our gratitude to Dr. Massimiliano Tamborrino, University of Warwick, and Yanzhi Chen, University of Cambridge, for their comments and suggestions that enhanced the quality of this manuscript.

\bibliographystyle{abbrvnat}
\bibliography{main}

\clearpage
\newpage

\appendix

\begin{center}
\LARGE
\textbf{Supplementary Material}
\end{center}
\normalsize

The supplementary material contains both additional methodological details on how to avoid singularities in the covariance matrix of the synthetic likelihood approximations, training specifications for the neural network, as well as further figures and an additional case-study.

\section{Particle degeneracy in Lookahead SIS and singularity of \texorpdfstring{$\tilde{\bs{\Sigma}}_{P, \bs{\theta}}$}{TEXT}}
\begin{figure}[t]
    \centering
    \includegraphics[scale=0.8]{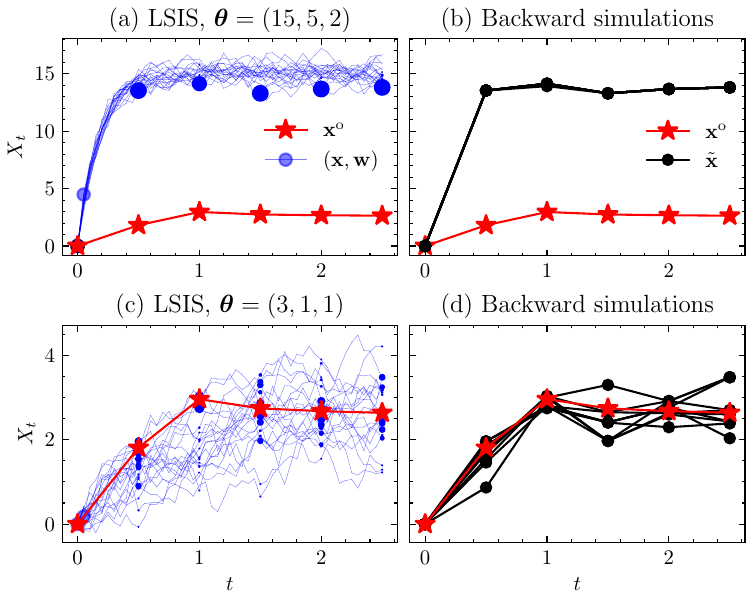}
    \caption{Illustration of a degenerate particle system and the corresponding backward trajectories. The observation (red) in both figures is a sample of an Ornstein--Uhlenbeck process with $\bs{\theta}^{\text{true}} = (3, 1, 1)$ and $\Delta t = 0.5$. In (a) the particle system obtained by running Lookahead SIS with $P = 20, A = 10$ and $\bs{\theta} = (15, 5, 2)$ is shown in blue. The parameter is deliberately chosen to be different from $\bs{\theta}^{\text{true}}$ and outside of the posterior region. For clarity, the normalized particle weights are depicted by the size of the circles only at the observational time instants. In (a) it can be seen that the weights accumulate on a single particle. In (b) $P = 20$ backward trajectories obtained by running the backward simulation particle smoother (on the particle system from (a)) are shown in black, and they all map onto a single trajectory. In (c), and similarly to (a), a particle system generated with $\bs{\theta} = \bs{\theta}^{\text{true}}$ is shown.  In (d) it can be seen that the corresponding backward trajectories are more variable, as compared to those in (b).}
    \label{fig:degen-system}
\end{figure}

When the particle system is degenerate, i.e. for every $k = 0, \ldots, N$, there exists a particle $j_k \in \{1, \ldots P \}$ such that $\omega_k^{j_k} \approx 1$, the particle approximation of the joint smoothing density (21) accumulates almost all of its mass into a single point $(\mathbf{x}_{1}^{j_1}, \ldots, \mathbf{x}_{N}^{j_N})$. The implications are that in the approximation of the synthetic likelihood ratio, the backward trajectories that are sampled in line 3 of Algorithm 4 are all the same, and therefore the covariance  $\tilde{\bs{\Sigma}}_{P, \bs{\theta}}$ of their summary statistics is singular. Particle degeneracy often occurs when the particles are ``far'' from the observation, because their weights, say at time $t_i$, are proportional to $\mathcal{N}(\bf{x}^{\rm o}_{{i}} \given \bf{x}^j_{t_{i} - h} + \mu(\bf{x}^j_{t_{i} - h}, \bs{\theta})h, \sigma^2(\bf{x}^j_{t_{i} - h}, \bs{\theta})h)$. To illustrate this, in Figure \ref{fig:degen-system} we show one particle system that is degenerate and one that is not, as well as the corresponding backward trajectories for both particle systems. 
Singular covariances are especially prevalent in the initial iterations of ABC-SMC, where the parameters are proposed from distributions that are very different from the posterior (e.g. the prior in the first iteration), and the acceptance thresholds are sufficiently large to ensure acceptance of said parameters. For a parameter-summary pair $(\bs{\theta}, \tilde{\bf{s}})$ for which $\tilde{\bs{\Sigma}}_{P, \bs{\theta}}$ is singular, we set $w(\bs{\theta}, \tilde{\bf{s}}) = 0$, because $\mathcal{N}(\tilde{\bf{s}} \given \tilde{\bs{\mu}}_{P, \bs{\theta}}, \tilde{\bs{\Sigma}}_{P, \bs{\theta}})$ cannot be evaluated. However, numerical instabilities can also arise when $\tilde{\bs{\Sigma}}_{P, \bs{\theta}}$ is near-singular, because $\mathcal{N}(\tilde{\bf{s}} \given \tilde{\bs{\mu}}_{P, \bs{\theta}}, \tilde{\bs{\Sigma}}_{P, \bs{\theta}}) \to \delta_{\tilde{\bs{\mu}}_{P, \bs{\theta}}}(\tilde{\bf{s}})$ when $\text{det}(\tilde{\bs{\Sigma}}_{P, \bs{\theta}}) \to 0$. To this end, we propose to use the condition number of $\tilde{\bs{\Sigma}}_{P, \bs{\theta}}$ as a measure of singularity, and set $w(\bs{\theta}, \tilde{\bf{s}})$ to zero if the condition number is bigger than a constant (in our experiments this constant is set to $10^3$). Formally, the condition number is defined as cond$(\tilde{\bs{\Sigma}}_{P, \bs{\theta}}) = ||\tilde{\bs{\Sigma}}_{P, \bs{\theta}}||_2 ||\tilde{\bs{\Sigma}}_{P, \bs{\theta}}^{-1}||_2$, and goes to $+\infty$ when $\tilde{\bs{\Sigma}}_{P, \bs{\theta}}$ is singular. While nullifying the weight for nearly singular covariance matrices enhances the stability of the synthetic likelihood ratios, it does not fully mitigate the issues relating to the high variance of the ratio. To further address this issue, we suggest 1) to compute the synthetic likelihood ratio on the log scale, and 2) to set the log SL ratio to $\log \text{SL}_i \coloneqq -\infty$ for a particle $i$ whose log SL ratio was greater than 0. This has the effect of limiting the influence of any individual particle with excessively large weight, on the weights normalization. Alternative approaches include temperature scaling, where weights are raised to a power to smooth out discrepancies. The choice of regularization technique depends on the specific characteristics and requirements of the problem at hand, and weight clipping is selected in this context for its simplicity and effectiveness in controlling the variance.
\section{CKLS with fixed $\gamma = 1/2$ (Cox-Ingersoll-Ross): inference for \texorpdfstring{$(\alpha, \beta, \sigma)$}{TEXT}} 

In this example, the CKLS model is considered with $\gamma$ fixed to \(\gamma = 1/2\), yielding the Cox--Ingersoll-Ross (CIR) process. A notable characteristic of the CIR process is that its transition densities are known. Specifically, the transition density from state \(x_s\) to state \(x_t\), for \(s < t\), is expressed as
\begin{equation}
    p(x_t \given x_s) = c e^{-u - v} (\frac{u}{v})^{q/2} I_q(2 \sqrt{uv}), \quad x, y \in \mathbb{R}_+, \label{eq:cir-tdensity}
\end{equation}
where
\begin{equation}
    c = \frac{2 \theta_2}{\theta_3^2 (1 - e^{-\theta_2 (t - s)})}, \quad q = \frac{2 \theta_1}{\theta_3^2} - 1, \quad u = c x_s e^{-\theta_2 (t - s)}, \quad v = c x_t.
\end{equation}
Here $I_q(\cdot)$ is the modified Bessel function of the first kind of order $q$. As data $\bf{x}^{\rm o}$, a discrete realization of the CIR model is generated with an initial state \(x_0 = 0\) and parameters \(\boldsymbol{\theta}_{\rm true} = (\alpha , \beta, \sigma) = (3, 2, 1)\) by sampling from the exact transition densities \eqref{eq:cir-tdensity} with \(\Delta t = 0.1\). For both the forward and data-conditional simulator, the number of subintervals between the observations is set to $A = 10$, hence the integration timestep is $h = \Delta t / A = 0.01$. The number of trajectories for the data-conditional simulator is set to $P = 50$. Despite the availability of the transition densities, to mimic an intractable scenario we choose as weighing functions the functions (26) with $\gamma = 1/2$. 

Next, we proceed to parameter inference for the CIR process, treating \(\gamma\) as known and fixed at \(\gamma = 1/2\), while \((\alpha, \beta, \sigma)\) are unknown. The CIR process is an example of a tractable diffusion because the likelihood function is available via a product of transition densities whose form is given by \eqref{eq:cir-tdensity}. Therefore exact inference via MCMC is possible. We define an independent prior distribution over the parameters \((\alpha, \beta, \sigma)\) by specifying \(\pi(\alpha, \beta, \sigma) = \mathcal{U}(\alpha \given 0, 20) \mathcal{U}(\beta \given 0, 10) \mathcal{U}(\sigma \given 0, 3)\). To obtain an initial estimate of the summary statistics, we pretrain PEN on 50,000 prior-predictive samples. The result of the analysis is in Figure 5(c) in the main paper, and is based on 10 independent runs on the same observed trajectory $\bf{x}^{\rm o}$. The figure displays the computation time versus the Wasserstein distance between the ABC posteriors and the reference posterior (obtained via MCMC). In this case, Figure 5(c) in the main paper clearly shows that with our sampler we require about 5 rounds  to localize the bulk of the posterior, and about 12-13 rounds with the standard sampler (also, notice that in this case we used a larger  prior distribution over $\sigma$, compared to previous experiments). In terms of running time, by comparing similar Wasserstein distances that are obtained at round 5 with our ABC-SMC-DC sampler, and at round 12 with the standard sampler, it takes about 2000 seconds to reach round 5 with ABC-SMC-DC and around 4100 seconds to reach round 12 with the standard ABC-SMC-F sampler, thus with our sampler we have a 2-fold acceleration on average. In Table 1 (see the main text) we show that by round 5 the data-conditional ABC-SMC achieves a significant reduction of the Wasserstein distance, while showing a larger acceptance rate than the pure forward approach. For later rounds, the acceptance rate of our simulator decreases, however this is expected since very small Wasserstein distances are achieved. 
\begin{figure}
    \centering
    \includegraphics[scale=0.93]{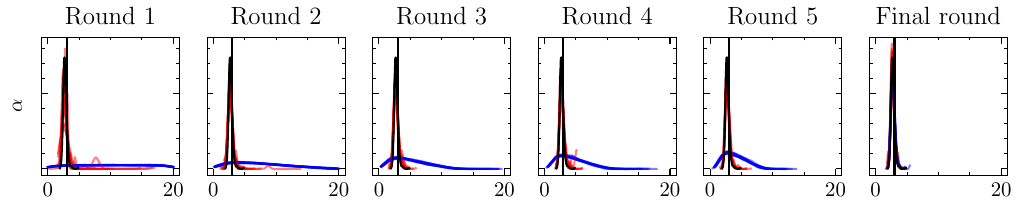}
    \includegraphics[scale=0.93]{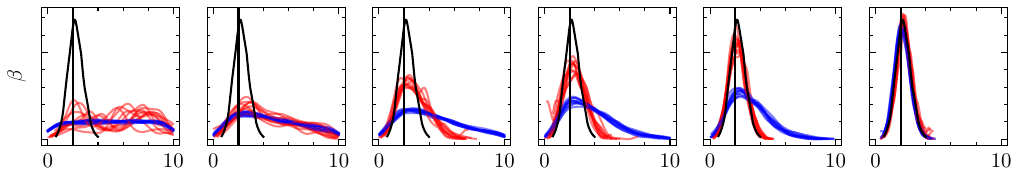}
    \includegraphics[scale=0.93]{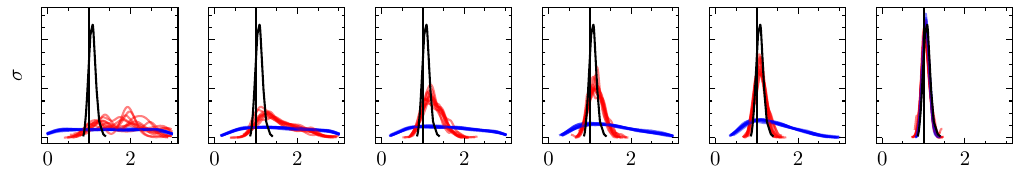}
    \caption{CIR model. Marginal posterior distributions across
several iterations (rounds)  for ABC-SMC-DC displayed in red, and ABC-SMC-F in blue. The reference marginal posteriors are depicted in black, and the true parameter values as black vertical lines.}
    \label{fig:cir-posteriors-row}
\end{figure}

\subsection{SDE with nonlinear drift}
\begin{figure}[tp]
    \centering
    \includegraphics[scale=0.93]{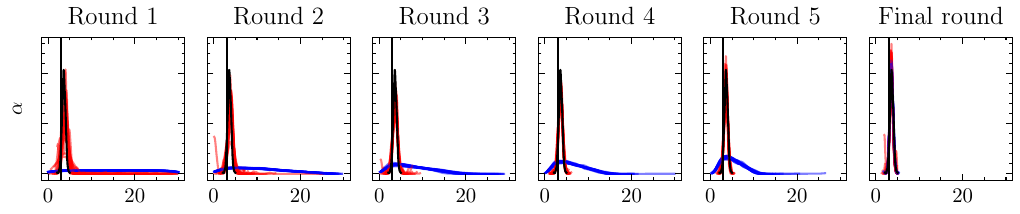}
    \includegraphics[scale=0.93]{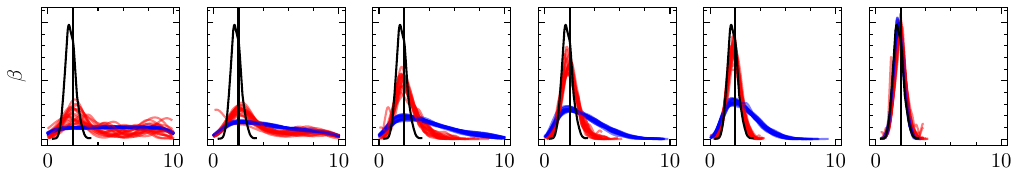}
    \includegraphics[scale=0.93]{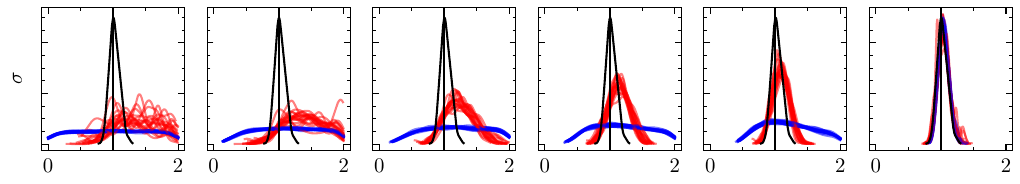}
    \caption{Nonlinear SDE model. Marginal posterior distributions for ABC-SMC-DC displayed in red, and ABC-SMC-F in blue. The reference marginal posteriors are depicted in black, and the true parameter values as black vertical lines.}
    \label{fig:nonlin-posteriors-row}
\end{figure}

This example considers the following SDE with nonlinear drift,
\begin{equation}
    \begin{cases}
            {\rm d}X_t = (\beta \alpha - \beta X_t + \sqrt{X_t}) {\rm d}t + \sigma \sqrt{X_t} {\rm d}B_t,  & \text{if } t > 0, \\
        X_0 = x_0, & \text{if } t = 0.
    \end{cases}
    \label{eq:nonlin-sde}
\end{equation}
for $\alpha, \beta \in \mathbb{R}$ and $\sigma \in \mathbb{R}_+$.
which has an intractable likelihood. We generate a discrete realization of \eqref{eq:nonlin-sde} with initial state $x_0 = 0.1$ and parameters $\boldsymbol{\theta}^{\rm true} = (\alpha, \beta, \sigma) = (3, 2, 1)$ by numerically solving the SDE via the EM method with integration timestep of $\Delta t = 0.0001$. After generating 100,000 data points we retained every 1,000th point and thus obtained a sample of length $n = 100$ at the timepoints $0, 0.1, \ldots, 10$. This set of $n$ observations is considered as our data $\bf{x}^o$. For the data-conditional simulator we set the number of trajectories to $P = 40$, and the number of subintervals between the observations to $A = 10$. We adopt an independent prior specification by taking $\pi(\boldsymbol{\theta}) = \mathcal{U}(\alpha \given 0, 30) \mathcal{U}(\beta \given 0, 10) \mathcal{U}(\sigma \given 0, 2)$, and pretrain PEN on 50,000 prior-predictive samples. The results of the analysis can be seen in Figure 5(d) of the main paper. We notice that our ABC-SMC-DC starts at a much lower (median) Wasserstein distance, and drops down to a stable value more rapidly than standard ABC-SMC, while showing a much larger acceptance rate than the standard forward approach (see Table 1 in the main paper). ABC-SMC-DC reaches the bulk of the posterior distribution at round 5 (1444 seconds) with a Wasserstein distance of $\approx 0.2$ (see also Figure \ref{fig:nonlin-posteriors-row}), whereas standard ABC-SMC reaches a similar Wasserstein distance at round 12 (4123 seconds), yielding a 2.85-fold improvement. 

\section{ABC posterior distributions for the Ornstein--Uhlenbeck study}
This section includes the ABC posterior distributions from the simulation study discussed in the main paper. See Figure \ref{fig:ou-posteriors-row} below.
\begin{figure}
    \centering
    \includegraphics[scale=0.93]{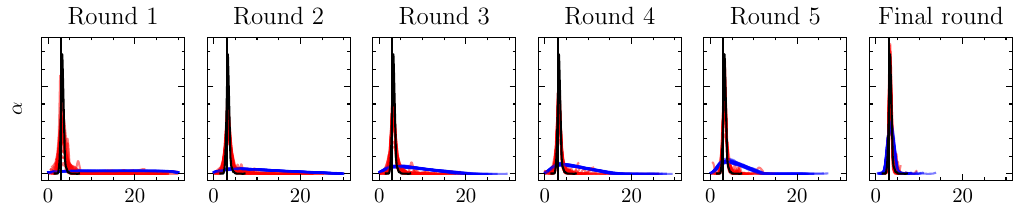}
    \includegraphics[scale=0.93]{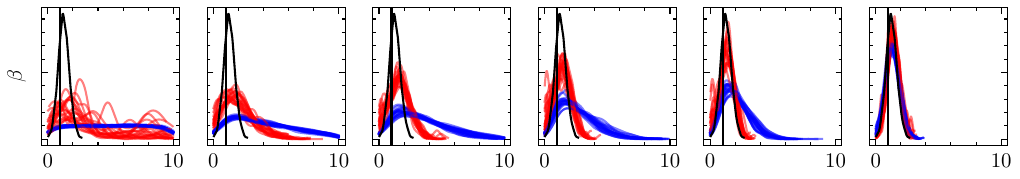}
    \includegraphics[scale=0.93]{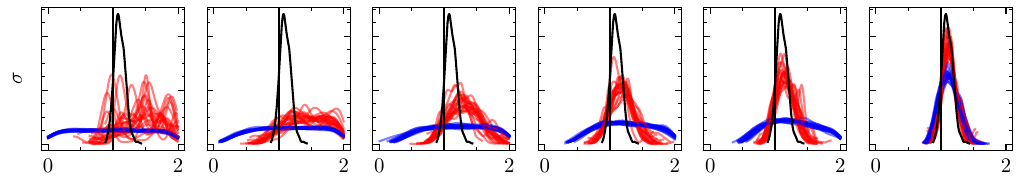}
    \caption{OU model example. Marginal posterior distributions across several iterations (rounds) of ABC-SMC: marginals from ABC-SMC-DC are in red, and those from the standard (forward) ABC-SMC-F are in blue. The reference marginal posteriors are depicted in black, and the true parameter values as black vertical lines.}
    \label{fig:ou-posteriors-row}
\end{figure}

\section{Trajectory plots for forward and data-conditional simulators}
\begin{figure}
    \centering\includegraphics[scale=0.7]{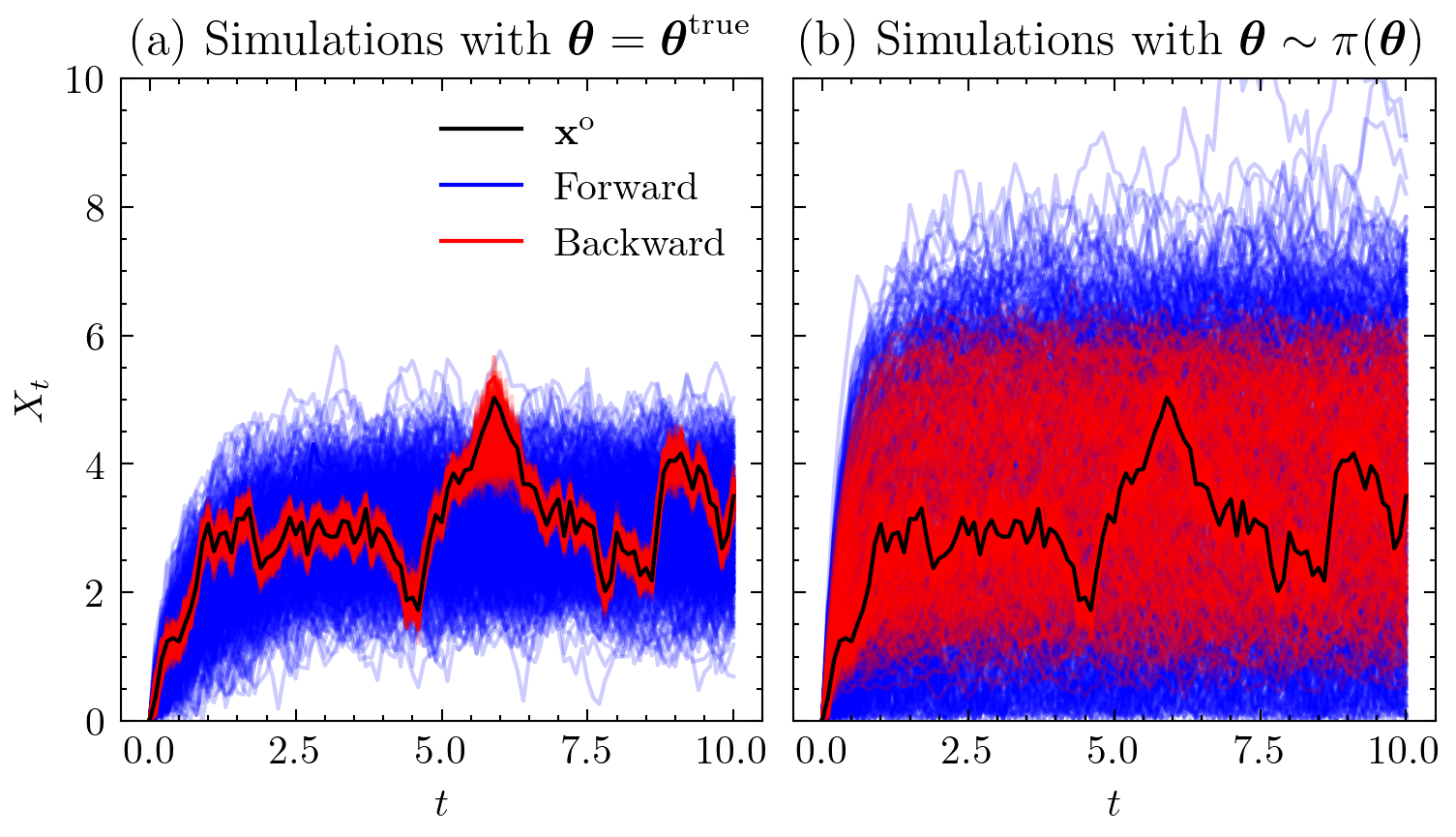}
    \centering\includegraphics[scale=0.7]{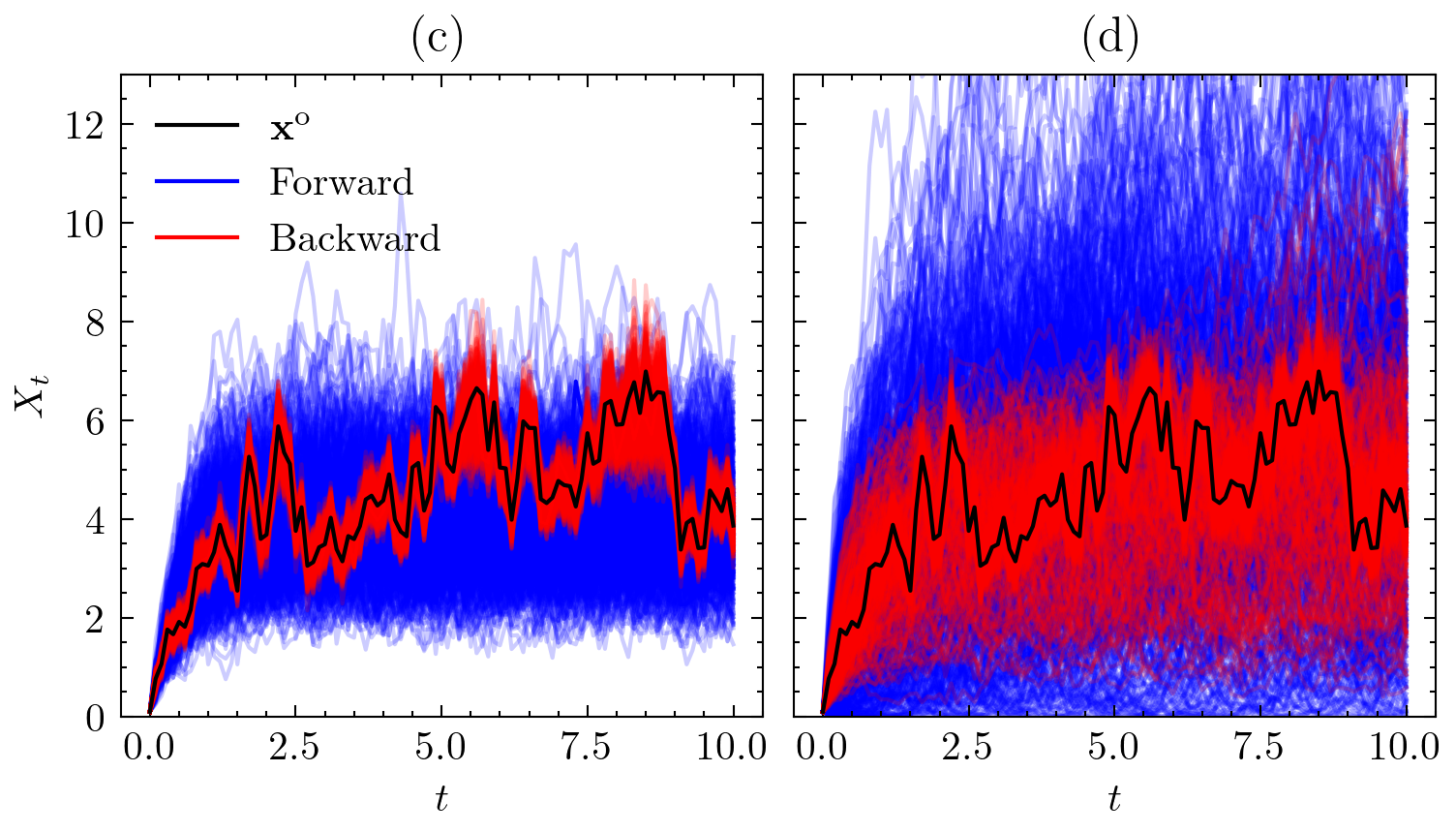}
    \centering\includegraphics[scale=0.7]{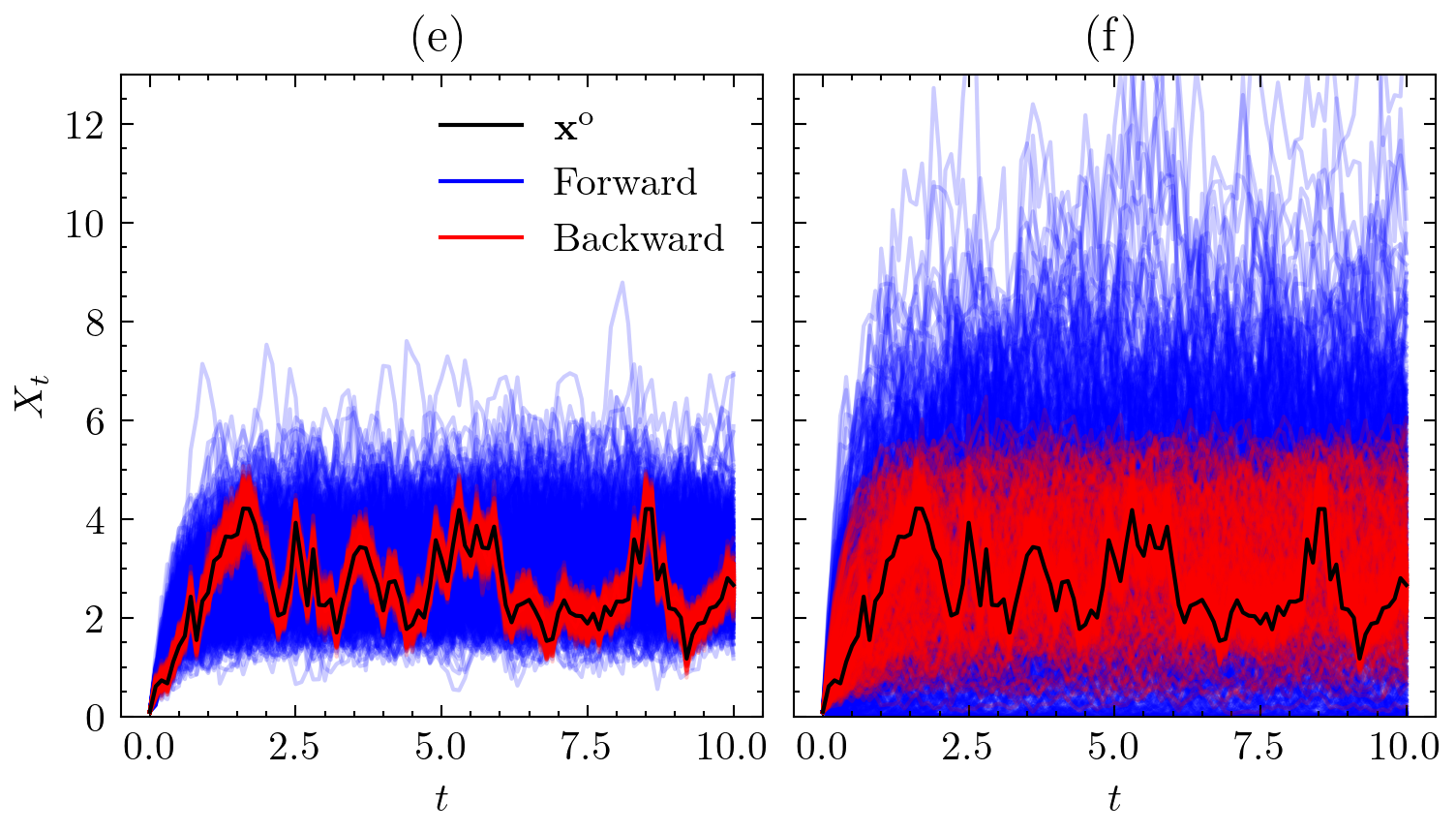}
    \caption{The observation is shown in black, the forward trajectories in blue, and the backward trajectories in red. The prior for all three examples is set to $\pi(\bs{\theta}) = \mathcal{U}(0, 7)\mathcal{U}(0, 3)\mathcal{U}(0.5, 1.5)$. Left column: simulations using parameters set at ground-truth values. Right colum: simulations using parameters sampled from their prior.
    Top row (a)-(b) Ornstein--Uhlenbeck SDE. Middle row (c)-(d) SDE with nonlinear drift. Bottom row (e)-(f) Cox--Ingersoll--Ross SDE.}
    \label{fig:trajectories-true-and-pps}
\end{figure}

We showcase the difference in simulated trajectories from both simulators. The observed data is the same as in the simulation study. The number of subintervals $A$ is set to 10 for both simulators, the number of particles for the data-conditional simulator is set to 30, and the weighing functions are the Gaussian densities induced by the EM scheme. A total of 500 trajectories are simulated per simulator, both when the parameter is set to the parameter truth, as well as when it is sampled from an arbitrary prior. The priors chosen in this subsection are different from those in the simulation study and only serve for illustration purposes. Trajectories for the OU, Nonlinear CIR and CIR SDEs can be seen in Figure \ref{fig:trajectories-true-and-pps}.

\begin{figure}
    \centering
    \includegraphics{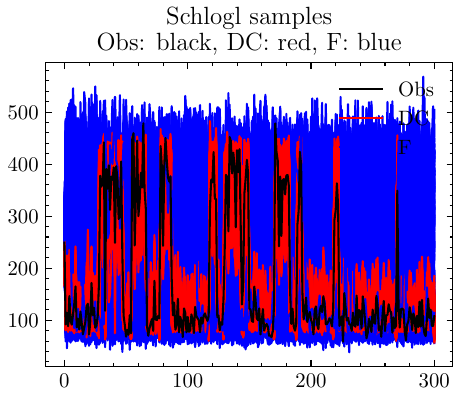}
    \caption{Sample paths from the Schl{\"o}gl model: data (black), data-conditional simulations (red), forward simulations (blue).}
    \label{fig:Schlogl}
\end{figure}

\section{Neural network and training specification}
For the CKLS family of models, PEN is structured as 2-100-100-100-dim($\bs{\theta}$) across all models. For the biochemical reaction networks, three 1D convolutional layers (25-50-100) are added prior to the the dense layers (100-100-100-dim($\bs{\theta}$)). PEN is trained for 1000 epochs, but if the validation loss remains stable for 200 epochs, training is halted. During each ABC-SMC iteration, a fresh dataset of $M$ samples is acquired. Again, 80\% of this is earmarked for training and the remaining 20\% for validation. These new chunks are added to the existing training and validation chunks. PEN is then retrained on the accumulated data for up to 1000 epochs, or until the validation loss stabilizes.

\section{Regular and irregular grids}
A regular observation grid, with equispaced intervals, is not a strict requirement for our method, but having an irregular observation grid would require a bit of attention. Naturally, if the observation grid is irregular, one would need to decide how many subintervals to include for every observational interval. However, there is a small caveat to this. We performed a new set of experiments on biochemical reaction networks (section 7.2), and in these experiments we assumed a regular observation grid with $\Delta = 1$, and we have set the forward integrator to integrate with a timestep of $h = \Delta / 100 = 0.01$. This means that in the Lookahead SIS algorithm, the Gaussian weighting function would need to be evaluated on the interval $[t - 0.01, t]$ for some observational time $t$. This turned out to be too aggressive and narrow, in that only a single particle would be given a very large weight. This is also a problem that shows up in the Bridge Particle Filter (BPF), albeit for different models. If the observation grid was not regular, then the aforementioned choice of weighting function would be too strict on some intervals, but not on others. In the BPF paper, the problem of a very narrow weighting function was solved by multiplying the covariance matrix (for a Gaussian weighting function) by  some constant. In our experiments for biochemical reaction networks we multiply the covariance matrix by $\Delta / 2$ instead of $h$. If the observation grid is not regular then the multiplication factor for every interval would need to be taken into account as well.

\end{document}